\documentclass[a4paper,11pt]{article}
\pdfoutput=1
\usepackage{jheppub}
\usepackage[utf8]{inputenc}

\newcommand{\ii}{\mathrm{i}}
\newcommand{\rme}{\mathrm{e}}

\newcommand{\vev}[1]{\langle #1 \rangle}
\newcommand{\Tr}{\mathrm{Tr}\,}
\newcommand{\tr}{\mathrm{tr}\,}
\newcommand{\diag}{\mathrm{diag}\,}

\newcommand{\one}{{\rm 1\kern -.9mm l}}

\newcommand{\be}{\begin{equation}}
\newcommand{\ee}{\end{equation}}
\newcommand{\p}{\partial}

\newcommand{\Lfour}{\Lambda}

\newcommand\undermat[2]{%
\makebox[0pt][l]{$\smash{\underbrace{\phantom{%
    \begin{matrix}#2\end{matrix}}}_{\text{$#1$}}}$}#2}
\newdimen\tableauside\tableauside=1.0ex
\newdimen\tableaurule\tableaurule=0.4pt
\newdimen\tableaustep
\def\phantomhrule#1{\hbox{\vbox to0pt{\hrule height\tableaurule
width#1\vss}}}
\def\phantomvrule#1{\vbox{\hbox to0pt{\vrule width\tableaurule
height#1\hss}}}
\def\sqr{\vbox{%
  \phantomhrule\tableaustep
\hbox{\phantomvrule\tableaustep\kern\tableaustep\phantomvrule\tableaustep}%
  \hbox{\vbox{\phantomhrule\tableauside}\kern-\tableaurule}}}
\def\squares#1{\hbox{\count0=#1\noindent\loop\sqr
  \advance\count0 by-1 \ifnum\count0>0\repeat}}
\def\tableau#1{\vcenter{\offinterlineskip
  \tableaustep=\tableauside\advance\tableaustep by-\tableaurule
  \kern\normallineskip\hbox
    {\kern\normallineskip\vbox
      {\gettableau#1 0 }%
     \kern\normallineskip\kern\tableaurule}%
  \kern\normallineskip\kern\tableaurule}}
\def\gettableau#1 {\ifnum#1=0\let\next=\null\else
  \squares{#1}\let\next=\gettableau\fi\next}
\newcommand{\Yfund}{\tableau{1}}
\newcommand{\Ysymm}{\tableau{2}}
\newcommand{\Yasymm}{\tableau{1 1}}

\def\XXint#1#2#3{{\setbox0=\hbox{$#1{#2#3}{\int}$}
     \vcenter{\hbox{$#2#3$}}\kern-.5\wd0}}

\usepackage{tikz}
\usetikzlibrary{graphs}
\usetikzlibrary{decorations.markings}

\tikzstyle{gauge} = [circle, text centered, draw=black, minimum height=1.5cm]
\tikzstyle{flavor} = [rectangle, text centered, draw=black, minimum height=1.5cm,minimum width=1.5cm]

\makeatletter
\pgfdeclareshape{barn}{
    \inheritsavedanchors[from={rectangle}]
    \inheritanchorborder[from={rectangle}] 
    \inheritanchor[from={rectangle}]{center}
    \inheritanchor[from={rectangle}]{south}
    \inheritanchor[from={rectangle}]{west}
    \inheritanchor[from={rectangle}]{east}
    \inheritbackgroundpath[from={rectangle}]
   
    \backgroundpath{
    \southwest \pgf@xa=\pgf@x \pgf@ya=\pgf@y
    \northeast \pgf@xb=\pgf@x \pgf@yb=\pgf@y
    \pgf@xc=\pgf@xb \advance\pgf@xc by-\pgf@xa     
    \pgf@yc=\pgf@yb \advance\pgf@yc by-\pgf@ya    
    \advance\pgf@yb by-0.5\pgf@yc
    \pgfpathmoveto{\pgfpoint{\pgf@xa}{\pgf@ya}}
    \pgfpathlineto{\pgfpoint{\pgf@xa}{\pgf@yb}}
    \pgfpatharc{180}{0}{.5\pgf@xc}  
    \pgfpathlineto{\pgfpoint{\pgf@xb}{\pgf@ya}}
    \pgfpathclose
 }
}
\makeatother

\tikzstyle{gaugedflavor} = [barn,draw, text centered, minimum height=1.5cm,minimum width=1.5cm,draw=black]

\title{\boldmath Surface operators, chiral rings and localization in $\mathcal{N}=2$ gauge theories}

\author[a,b]{S.~K.~Ashok,}

\affiliation[a]{Institute of Mathematical Sciences \\
   C.\,I.\,T.~Campus, Taramani\\
   Chennai, India 600113\\}
   
\affiliation[b]{Homi Bhabha National Institute\\
Training School Complex, Anushakti Nagar,\\ 
Mumbai, India 400085\\}
\emailAdd{sashok@imsc.res.in}

\author[c,d]{M.~Bill\`o,}
\affiliation[c]{Universit\`a di Torino, Dipartimento di Fisica}

\affiliation[d]{Arnold-Regge Center and I.\,N.\,F.\,N. - sezione di Torino, \\
Via P. Giuria 1, I-10125 Torino, Italy\\}
\emailAdd{billo@to.infn.it}

\author[c]{E.~Dell'Aquila,}
\emailAdd{edellaquila@gmail.com}

\author[c,d]{M.~Frau,}
\emailAdd{frau@to.infn.it}

\author[a,b]{V.~Gupta,}
\emailAdd{varungupta@imsc.res.in}

\author[a,b]{R.~R.~John,}
\emailAdd{renjan@imsc.res.in}

\author[e,d]{and A.~Lerda\,}
\affiliation[e]{Universit\`a del Piemonte Orientale, Dipartimento di Scienze e Innovazione Tecnologica\\
Viale T. Michel 11, I-15121 Alessandria, Italy\\}
\emailAdd{lerda@to.infn.it}

\abstract{
We study half-BPS surface operators in supersymmetric gauge theories in four and five dimensions 
following two different approaches. 
In the first approach we analyze the chiral ring equations for certain quiver 
theories in two and three dimensions, coupled respectively to four- and five-dimensional gauge theories. 
The chiral ring equations, which arise from extremizing a twisted chiral superpotential, are solved as 
power series in the infrared scales of the quiver theories. 
In the second approach we use equivariant localization and obtain the twisted chiral superpotential 
as a function of the Coulomb moduli of the four- and five-dimensional gauge theories, and find a perfect 
match with the results obtained from the chiral ring equations.
In the five-dimensional case this match is achieved after solving a number of subtleties in the 
localization formulas which amounts to choosing a particular residue prescription in the integrals that yield 
the Nekrasov-like partition functions for ramified instantons. We also comment on the necessity 
of including Chern-Simons terms in order to match the superpotentials obtained from
dual quiver descriptions of a given surface operator.
}

\keywords{$\mathcal{N}=2$ gauge theories, instantons, surface operators}

\preprint{ARC-17-5}

\begin{document}
\maketitle
\flushbottom
\section{Introduction}

In this paper we study the low-energy effective action that governs the dynamics of 
half-BPS surface operators in theories with eight supercharges. We focus
on pure SU($N$) theories in four dimensions and in five dimensions compactified on a circle,
and explore their Coulomb branch where the adjoint scalars acquire a vacuum expectation value (vev). 

In four dimensions, a surface defect supports on its world-volume a two-dimensional gauge theory
that is coupled to the ``bulk" four-dimensional theory, see \cite{Gukov:2014gja} for a review. 
This combined 2d/4d system is described by two holomorphic functions: the prepotential ${F}$ 
and the twisted superpotential ${W}$.
The prepotential governs the dynamics of the bulk theory and depends on 
the Coulomb vev's and the infra-red (IR) scale of the gauge theory in four dimensions. 
The twisted superpotential controls the two-dimensional dynamics on the surface operator, 
and is a function of the continuous parameters labeling the defect, the two-dimensional IR scales, and 
also of the Coulomb vev's and the strong-coupling scale of the bulk gauge theory. 
The twisted superpotential thus describes the coupled 2d/4d system.

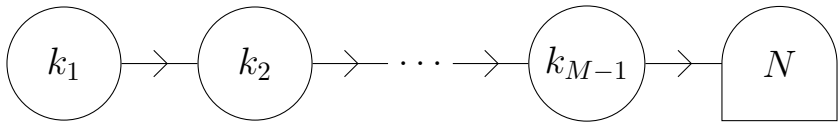
\begin{figure}[ht]
\begin{center}
\begin{tikzpicture}[decoration={markings, mark=at position 0.6 with {\draw (-5pt,-5pt) -- (0pt,0pt);       
         \draw (-5pt,5pt) -- (0pt,0pt);}}]
  \matrix[row sep=10mm,column sep=5mm] {
      \node (g1)[gauge] {\Large $k_1$};  & & \node(g2)[gauge] {\Large $k_2$};  && \node(dots){\Large 
      $\ldots$};
      & & \node(glast)[gauge] {\Large $k_{M-1}$};& &\node(gfN)[gaugedflavor]{\Large $N$};\\
  };
\graph{(g1) --[postaction={decorate}](g2) --[postaction={decorate}](dots)--[postaction={decorate}](glast) 
--[postaction={decorate}](gfN);};
\end{tikzpicture}
\end{center}
\vspace{-0.5cm}
\caption{The quiver which describes the generic surface operator in pure SU$(N)$ gauge theory.}
\label{quiverpicgeneric1}
\end{figure}

Surface operators in theories with eight supercharges can be studied from diverse points of view. 
One approach is to treat them as monodromy defects (also known as Gukov-Witten defects)
in four dimensions along the lines discussed 
in \cite{Gukov:2006jk, Gukov:2008sn} and compute the corresponding twisted superpotential ${W}$ 
using equivariant localization as shown for example 
in \cite{Kanno:2011fw, Nawata:2014nca, Ashok:2017odt, Gorsky:2017hro}. 
A second approach is to focus on the two-dimensional world-volume theory on the surface 
operator \cite{Gaiotto:2009fs}. In the superconformal theories of class ${\mathcal S}$, a microscopic 
description of a generic co-dimension 4 surface operator in terms of $(2,2)$ supersymmetric quiver gauge 
theories in two dimensions was realized in \cite{Gomis:2014eya}. Here we focus on a generic 
co-dimension 2 surface operator in 
pure ${\mathcal N}=2$ SU$(N)$ gauge theory \cite{Gaiotto:2013sma}, 
which has a microscopic description as a quiver gauge theory of the type shown 
in Fig.~\ref{quiverpicgeneric1}.

Here the round nodes, labeled by an index $I$, correspond to U($k_I$) gauge theories in two dimensions
whose field strength is described by a twisted chiral field $\Sigma^{(I)}$.
The rightmost node represents the four-dimensional ${\mathcal N}=2$ gauge theory whose SU($N$) 
gauge group acts as a flavor group for the last two-dimensional node. The arrows correspond to 
(bi-)fundamental matter multiplets that are generically massive. Integrating out these fields 
leads to an effective action for the twisted chiral fields which, because of the two-dimensional 
$(2,2)$-supersymmetry, is encoded in a twisted chiral superpotential $W$. 
The contribution to $W$ coming from the massive fields attached to the last node depends 
on the four-dimensional dynamics of the SU($N$) theory and in particular on its resolvent 
\cite{Gaiotto:2013sma}. 
In this approach a key role is played by the twisted
chiral ring equations that follow from extremizing the twisted superpotential with 
respect to $\Sigma^{(I)}$. The main idea is that by evaluating ${W}$ on the solutions to the twisted chiral 
ring equations one should reproduce precisely the superpotential calculated using localization. 

In this work, we extend this analysis in a few directions. We show that there exists a precise 
correspondence between the choice of massive vacua in two dimensions and the Gukov-Witten defects of 
the SU($N$) gauge theory labeled by the partition $[n_1, \ldots n_M]$ with $n_1+\cdots +n_M=N$. 
We also describe the relation between the $(M-1)$ dynamically generated scales $\Lambda_I$ associated to the Fayet-Iliopoulos (FI) parameters for the two-dimensional nodes 
and the $(M-1)$ dimensionful parameters that naturally occur in the ramified instanton counting problem. 
Both the chiral ring equations and the localization methods can be extended to five-dimensional theories
compactified on a circle of circumference $\beta$, {\it{i.e.}} to theories defined 
on $\mathbb{R}^4\times S^1_{\beta}$. In this case, surface operators correspond
to codimension-2 defects wrapping $S^1_{\beta}$ and supporting a three-dimensional gauge theory
coupled to the bulk five-dimensional theory. In the 3d/5d case, one again has 
a quiver theory, and as before its infrared dynamics is encoded in a twisted chiral superpotential.
However, the form of the superpotential is modified to take into account the presence of a
compactified direction. The twisted chiral rings for purely
three-dimensional quiver theories have been studied in great detail in \cite{Gaiotto:2013bwa}. 
Here we extend this analysis and propose that the coupling between the last three-dimensional 
gauge node and the compactified five-dimensional theory is once again determined via the resolvent of the
latter.
With this assumption, the analysis of the modified twisted chiral ring equations as well as the 
choice of vacuum follow exactly the same pattern as in the 2d/4d case. An important and non-trivial 
check of this proposal is provided by the perfect agreement between the twisted superpotential 
obtained from solving the chiral ring equations and the one obtained from localization in five dimensions, 
which we perform in several examples. 

In the 2d/4d case, the quiver theory on the defect can be mapped to 
other quiver theories by chains of Seiberg-like dualities, which lead to different quiver realizations
of the same Gukov-Witten defect \cite{Benini:2014mia,Closset:2015rna,Gorsky:2017hro}. We show that, with an appropriate ansatz, the solutions of the twisted 
chiral ring equations for such dual theories lead to the same twisted superpotential.
We obtain strong indications that each such superpotential matches the result of a 
localization computation carried out with a different residue prescription. If we 
promote the quiver theories to the compactified 3d/5d set-up, 
the superpotentials still agree at the classical level but, in general, they differ when quantum corrections are 
taken into account.  The 3d/5d quiver gauge theories
can be extended to include Chern-Simons (CS) terms in their effective action.
Quite remarkably, we find in a simple but significant 
example that the equivalence between the dual quiver realizations of the same defect is restored at 
the quantum level if suitably chosen CS terms are added to the superpotentials.

The paper proceeds as follows. In Section~\ref{quiver2d4d}, we study the coupled 2d/4d system and solve 
the twisted chiral ring equations as power series in the IR scales of the theory. In 
Section~\ref{quiver3d5d}, we lift the discussion to coupled 3d/5d systems compactified on a circle. 
In Section~\ref{localize4d5d}, we analyze the ramified instanton counting in four and five dimensions 
and show that the effective twisted chiral superpotential calculated using localization methods exactly 
matches the one obtained from the solution of the chiral ring equations in the earlier sections. 
In Section~\ref{secn:dual}, we discuss the relation between different quiver realizations of the same
surface defect, and show that the equivalence between two dual realizations, 
which is manifest in the 2d/4d case, is in general no longer true in the 3d/5d case. 
We also show in a specific example that the duality is restored by adding suitable Chern-Simons terms.
Finally, in Section~\ref{conclusions} we present our conclusions and discuss some possible extensions 
of our results. Some technical details are collected in the appendices.

\section{Twisted superpotential for coupled 2d/4d theories}
\label{quiver2d4d}

In this section our focus is the calculation of the low-energy effective action for surface operators 
in pure ${\mathcal N}=2$ SU$(N)$ supersymmetric gauge theories in four dimensions. 
As mentioned in the Introduction,
surface operators can be efficiently described by means of a coupled 2d/4d system in which 
the two-dimensional part is a $(2,2)$-supersymmetric quiver gauge theory with (bi-)fundamental 
matter, as shown in Fig.~\ref{quiverpicgeneric1}. 
Such coupled 2d/4d systems have an alternative description as Gukov-Witten monodromy 
defects \cite{Gukov:2006jk, Gukov:2008sn}. The discrete data that label these defects 
correspond to the partitions of $N$, and can be summarized in the 
notation $\mathrm{SU}(N)[n_1,\ldots, n_M]$ where $n_1+\cdots+n_M=N$. The $M$ integers $n_I$ are
related to the breaking pattern (or Levi decomposition) of the four-dimensional gauge group on the defect, 
namely
\begin{equation}
\mathrm{SU}(N)~~\longrightarrow~~\mathrm{S}\big[\mathrm{U}(n_1)\times\ldots\times\mathrm{U}(n_M)\big]~.
\label{Levi}
\end{equation}
They also determine the ranks $k_I$ of the two-dimensional gauge groups of the quiver 
in Fig.~\ref{quiverpicgeneric1} according to
\begin{equation}
k_I= n_1+\cdots+n_{I}~.
\label{kI}
\end{equation}
The (bi-)fundamental fields connecting two nodes turn out to be massive. Integrating them out leads to 
the low-energy effective action for the gauge multiplet. In the $I$-th node the gauge multiplet is described by a 
twisted chiral field $\Sigma^{(I)}$ and the low-energy effective action
is encoded in a twisted chiral superpotential ${W}(\Sigma^{(I)})$. 
The vacuum structure can be determined by the twisted chiral ring equations, which 
take the form \cite{Nekrasov:2009uh, Nekrasov:2009ui, Nekrasov:2009rc}
\begin{equation}
\exp\Big(\frac{\partial{W}}{\partial\Sigma^{(I)}_s} \Big) = 1
\label{vaccumeq}
\end{equation}
where $\Sigma^{(I)}_s$ are the diagonal components, with $s=1,\ldots,k$ \cite{Hanany:1997vm}.
This exponentiated form of the equations is a consequence of the electric fluxes which can be 
added to minimize the potential energy and which lead to linear (in $\Sigma^{(I)}$) terms in 
the effective superpotential. 

We extend this analysis in the following manner: first of all, we show that in the 
classical limit there is a very specific choice of solutions to the twisted chiral ring equations 
that allows us to make contact with the twisted chiral superpotential calculated using localization. 
We establish the correspondence between the continuous parameters labeling the monodromy 
defect and the dynamically generated scales of the two-dimensional quiver 
theory. We then show that quantum corrections in 
the quiver gauge theory are mapped directly to corrections in the twisted superpotential due 
to ramified instantons of the four-dimensional theory. 

\subsection{SU(2)}
\label{SU2}
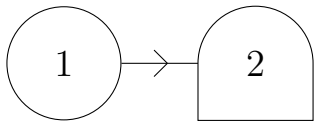
\begin{figure}[ht]
\begin{center}
\begin{tikzpicture}[decoration={markings, mark=at position 0.6 with {\draw (-5pt,-5pt) -- (0pt,0pt);
                \draw (-5pt,5pt) -- (0pt,0pt);}}]
  \matrix[row sep=10mm,column sep=5mm] {
      \node(g1)[gauge] {\Large $1$};      & &\node(gf2)[gaugedflavor]{\Large $2$};\\
  };
\graph{(g1) --[postaction={decorate}](gf2);};
\end{tikzpicture}
\end{center}
\vspace{-0.5cm}
\caption{The quiver representation of the SU(2)[1,1] surface operator.}
\label{quiversu2}
\end{figure}

As an illustrative example, we consider the simple surface operator in the 
pure SU(2) theory which is 
represented by the partition $[1,1]$. {From} the two-dimensional perspective,
the effective dynamics is described by a non-linear sigma model with target-space $\mathbb{CP}^1$, coupled
to the four-dimensional SU(2) gauge theory in a particular way that we now describe. We use the gauged linear sigma model 
(GLSM) description of this theory \cite{Witten:1993yc,Hanany:1997vm} 
in order to study its vacuum structure. We essentially follow the discussion in \cite{Gaiotto:2013sma} 
although, as we shall see in detail, there are some differences in our analysis. 
The GLSM is a U(1) gauge theory with two chiral multiplets in the fundamental representation, 
that can be associated to the quiver drawn in  Fig.~\ref{quiversu2}.

Let us first analyze the simple case in which the quantum effects of the SU(2) theory are neglected. 
We consider a generic point in the Coulomb branch parameterized by the vev's $a_1=-a_2=a$ of the
adjoint SU(2) scalar field $\Phi$ in the vector multiplet. These have the interpretation of twisted masses for 
the chiral multiplet of the two-dimensional U(1) gauge theory.
The theory obtained by integrating out this massive multiplet has been studied in some detail
in a number of works and here we merely present the resulting effective action which takes the simple
form\,\footnote{For notational simplicity we denote the superfield $\Sigma$ by its lowest scalar
component $\sigma$.}:
\begin{equation}
\begin{aligned}
{W}&= 2\pi\ii\,\tau(\mu)\,\sigma -\sum_{i=1}^2 (\sigma-a_i) \Big(\log
\frac{\sigma - a_i}{\mu} - 1\Big)\\
&=2\pi\ii\,\tau(\mu)\,\sigma -
\Tr\bigg[ (\sigma-\Phi) \Big(\log\frac{\sigma - \Phi}{\mu} - 1\Big) \bigg] ~.
\end{aligned}
\end{equation}
Here $\mu$ is the ultra-violet (UV) cut-off which we eventually take to infinity, and $\tau(\mu)$ is the bare FI parameter at the scale $\mu$. We can rewrite this superpotential using another scale $\mu^\prime$ and get
\begin{equation}
{W} = \Big(2\pi\ii\,\tau(\mu)-2\log\frac{\mu^\prime}{\mu}\,\Big)
\sigma - \Tr\bigg[ (\sigma-\Phi) \Big(\log\frac{\sigma - \Phi}{\mu^\prime} - 1\Big) \bigg]~.
\label{tildeW}
\end{equation}
{From} the coefficient of the linear term in $\sigma$, we identify the 
running of the FI coupling%
\footnote{\label{foot:rrg}Recall that $\tau$ is actually the complexification of the FI parameter $r$ with 
the $\theta$-angle: $2\pi\ii\tau = \ii\theta - r$. The sign of the coefficient of the logarithmic running (\ref{tildeW}) is such that $r(\mu^\prime)$ grows with the scale $\mu^\prime$. The same is true in the other 
cases we consider.}:
\begin{equation}
2\pi\ii\, \tau(\mu^\prime) =2\pi\ii \,\tau(\mu)-2\log\frac{\mu^\prime}{\mu}~.
\end{equation}
In particular, we can choose to use the complexified 
IR scale $\Lambda_1$ at which $\tau(\Lambda_1)=0$, so that 
\begin{equation}
{W} =  - \Tr\bigg[ (\sigma-\Phi) \Big(\log\frac{\sigma - \Phi}{\Lambda_1} - 1\Big) \bigg]~.
\label{tildeWL1}
\end{equation}
In this way we trade the UV coupling $\tau(\mu)$ for the dynamically generated scale $\Lambda_1$.   

Let us now turn on the dynamics of the four-dimensional SU(2) gauge theory. As pointed out in
\cite{Gaiotto:2013sma}, this corresponds to considering the following superpotential:
\begin{equation}
{W}= - \left\langle \Tr\left[ (\sigma-\Phi)
\Big(\log\frac{\sigma - \Phi}{\Lambda_1} - 1\Big) \right]\right\rangle~.
\label{quantumW}
\end{equation}
The angular brackets signify taking the quantum corrected vev of the chiral observable in 
the four-dimensional SU(2) theory. 
The twisted chiral ring equation is obtained by extremizing ${W}$ and is given by
\begin{equation}
\exp\Big(\frac{\partial{W}}{\partial \sigma} \Big) = 1~,
\end{equation}
which, using the superpotential (\ref{quantumW}), is equivalent to
\begin{equation}
\exp \left\langle \Tr\log\frac{\sigma- \Phi}{\Lambda_1}\right\rangle =  1~.
\label{creq1}
\end{equation}
As explained in \cite{Gaiotto:2013sma}, the left-hand side of (\ref{creq1}) is simply the 
integral of the resolvent of the pure ${\mathcal N}=2$ SU(2) theory in four dimensions which 
takes the form \cite{Cachazo:2002ry}:
\begin{equation}
\left\langle\Tr\log\frac{\sigma- \Phi}{\Lambda_1}\right\rangle = \log\left(\frac{P_2(\sigma) 
+ \sqrt{P_2(\sigma)^2 - 4\Lambda^4}}{2\Lambda_1^2}\right) ~.
\label{trlog}
\end{equation}
Here $\Lambda$ is the four-dimensional strong coupling scale of the SU(2) theory and 
\begin{equation}
P_2(\sigma)=\sigma^2-u
\end{equation}
is the characteristic polynomial appearing in the Seiberg-Witten solution where
\begin{equation}
u=\frac{1}{2}\,\big\langle\Tr\Phi^2\big\rangle = a^2+\frac{\Lfour^4}{2 a^2}+ \frac{5 \Lfour^8}{32 a^6}+\ldots
\label{u}
\end{equation}
Using (\ref{trlog}) and performing some simple manipulations, 
we find that the twisted chiral ring relation (\ref{creq1}) becomes
\begin{equation}
P_2(\sigma)= \Lambda_1^2+ \frac{\Lfour^4}{\Lambda_1^2}
\end{equation}
from which we obtain the two solutions
\begin{equation}
\sigma^{\pm}_\star(u,\Lambda_1)= \pm \,\sqrt{u +  \Lambda_1^2+ \frac{\Lfour^4}{\Lambda_1^2}}~.
\label{SU2solns}
\end{equation}
Notice the explicit presence of two different scales,
$\Lambda_1$ and $\Lambda$, which are related respectively to the two-dimensional and the
four-dimensional dynamics. Clearly, the purely two-dimensional result can be recovered by taking 
the $\Lambda\rightarrow 0$ limit. 
We can now substitute either one of the solutions of the chiral ring equation into the twisted 
chiral superpotential and obtain a function ${W}^{\pm}_\star$. 
The proposal in \cite{Gaiotto:2013sma} is that this should reproduce the twisted superpotential 
calculated using localization methods. We shall explicitly verify this in Section~\ref{localize4d5d}, but here 
we would like to point out an important simplification that occurs in this calculation.

Let us consider the twisted effective superpotential evaluated on the $\sigma^+_\star$ solution 
of the chiral ring relations, namely 
\begin{equation}
W^{+}_\star\big(u,\Lambda_1\big)\,\equiv\,
W\big(\sigma^+_\star(u,\Lambda_1),\Lambda_1\big) ~.
\label{Wsplit}
\end{equation}
While ${W}^+_\star$ itself is complicated, its logarithmic derivative with respect to $\Lambda_1$
takes a remarkably simple form. In fact $W^+_\star$ seems to depend on $\Lambda_1$ both explicitly and through the solution $\sigma^+_\star$, but on shell $\partial W/\partial\sigma\big|_{\sigma_\star^+}=0$ 
and so we simply have
\begin{equation}
\label{dWdL}
\Lambda_1 \frac{d W^+_\star}{d\Lambda_1} = \Lambda_1 \frac{\partial W}{\partial\Lambda_1}\bigg|_{\sigma_\star^+}
= 2\sigma_\star^+~.
\end{equation}
where in the last step we used (\ref{quantumW}) and took into account the tracelessness of $\Phi$.

Using the explicit form of the solution given in (\ref{SU2solns}), and inserting in it the weak-coupling 
expansion (\ref{u}) of $u$, we thus obtain
\begin{equation}
\begin{aligned}
\frac{1}{2} \,\Lambda_1\frac{d{W}^{+}_\star}{d \Lambda_1}
= a+ \frac{1}{2a}\Big(\Lambda_1^2 &\,+ \frac{\Lfour^4}{\Lambda_1^2} \Big) 
-\frac{1}{8a^3}\Big(\Lambda_1^4 + \frac{\Lfour^8}{\Lambda_1^4}\Big) \\
&+\frac{1}{16a^5}\Big(\Lambda_1^6 +\Lambda_1^2\Lfour^4+ \frac{\Lfour^8}{\Lambda_1^2}
+ \frac{\Lfour^{12}}{\Lambda_1^6}\Big) +\ldots~.
\end{aligned}
\label{Wsu2TCR}
\end{equation}
As we shall show later in Section~\ref{localize4d5d}, this result precisely matches the derivative of the twisted 
effective superpotential calculated using localization for the simple surface operator in the SU$(2)$ gauge 
theory, provided we suitably relate the dynamically generated scale $\Lambda_1$ of the two-dimensional
theory to the ramified instanton counting parameter in presence of the monodromy defect. 

\subsection{Twisted chiral ring in quiver gauge theories}
We will now show that the procedure described above generalizes in a rather simple way to any
surface operator in the SU$(N)$ gauge theory labeled by a partition of $N$. 
In this case, however, it will not be possible to solve exactly the twisted chiral ring equations as we did 
in the SU$(2)$ theory. We will have to develop a systematic perturbative approach
in order to obtain a semi-classical expansion for the twisted chiral superpotential around a particular classical
vacuum. Proceeding in this way we again find that the derivatives of the twisted superpotential 
with respect to the various scales have simple expressions in terms of combinations of the twisted 
chiral field $\sigma$ evaluated in the appropriately chosen vacuum. 

Following \cite{Gaiotto:2013sma}, we consider a quiver gauge theory of the form
\begin{equation}
\mathrm{U}(k_1) \times \mathrm{U}(k_2) \times \ldots \times \mathrm{U}(k_{M-1})
\label{Uks}
\end{equation}
with (bi)-fundamental matter between successive nodes, coupled to a 
pure $\mathcal{N}=2$ theory in four dimensions with gauge group SU($N$) acting as a
flavor symmetry for the rightmost factor in (\ref{Uks}). All this is represented in Fig.~\ref{quiverpicgeneric1}.
We choose an ordering such that
\begin{equation}
k_1 < k_2 < k_3 \ldots < k_{M-1} < N~,
\label{ordering}
\end{equation}
where the $k_I$'s are related to the entries of the partition of $N$ labeling the surface operator as indicated in
(\ref{kI}).
Our first goal is to obtain the twisted chiral ring of this 2d/4d system. Only the diagonal
components of $\sigma$ are relevant for this purpose \cite{Hanany:1997vm}, and thus for the $I$-th gauge group we take
\begin{equation}
\sigma^{(I)} = \mathrm{diag}\big(\sigma^{(I)}_1,\sigma^{(I)}_2,\ldots,\sigma^{(I)}_{k_I}\big)~.
\label{sigmadiag}
\end{equation}
The (bi)-fundamental matter fields are massive and their (twisted) mass is proportional to the difference 
in the expectation values of the $\sigma$'s in the two nodes connected by the matter multiplet. 
In order to minimize the potential energy, the twisted chiral field $\sigma^{(I)}$ gets a vev 
and this in turn leads to a non-vanishing mass for the (bi)-fundamental matter. 
Integrating out these massive fields, we obtain the following effective superpotential 
\begin{equation}
\begin{aligned}
{W} =& ~2\pi\ii \sum_{I=1}^{M-1}\sum_{s=1}^{k_I} \tau_I(\mu)\,\sigma^{(I)}_s-\!
\sum_{I=1}^{M-2} \sum_{s=1}^{k_{I}} \sum_{t=1}^{k_{I+1}}
\varpi\big(\sigma^{(I)}_s - \sigma^{(I+1)}_t\big) \\
&\qquad-\!\sum_{s=1}^{k_{M-1}}\!\Big\langle 
\Tr \varpi\big(\sigma^{(M-1)}_s - \Phi\big) \Big\rangle
\end{aligned}
\label{tildeW1}
\end{equation}
where, for compactness, we have introduced the function
\begin{equation}
\label{defvarpi}
\varpi(x) = x\,\Big(\log \frac{x}{\mu} -1\Big)
\end{equation}
with $\mu$ being the UV cut-off scale. 
Similarly to the SU$(2)$ example previously considered, also here we can trade 
the UV parameters $\tau_I(\mu)$ for the dynamically generated scales  $\Lambda_I$ for each of
the gauge groups in the quiver. To this aim, we unpackage of the terms containing the
$\varpi$-function and rewrite them as follows:
\begin{equation}
\begin{aligned}
\varpi\big(\sigma^{(I)}_s - \sigma^{(I+1)}_t\big) =
&\,\sigma^{(I)}_s\bigg(\log\frac{\sigma^{(I)}_s-\sigma^{(I+1)}_t}{\Lambda_I}-1\bigg)-
\sigma^{(I+1)}_t\bigg(\log\frac{\sigma^{(I)}_s-\sigma^{(I+1)}_t}{\Lambda_{I+1}}-1\bigg)\\
&+\sigma^{(I)}_s \log\frac{\Lambda_I}{\mu}-\sigma^{(I+1)}_t \log\frac{\Lambda_{I+1}}{\mu}
\end{aligned}
\end{equation}
for $I=1,\ldots, M-2$, and
\begin{equation}
\begin{aligned}
\Tr \varpi\big(\sigma^{(M-1)}_s - \Phi\big)=&\,\Tr \left[(\sigma^{(M-1)}_s - 
\Phi)\bigg(\log\frac{\sigma^{(M-1)}_s - \Phi}{\Lambda_{M-1}}-1\bigg)\right]\\
&+N\,\sigma^{(M-1)}_s \log\frac{\Lambda_{M-1}}{\mu}~.
\end{aligned}
\end{equation}
Considering the linear terms in the $\sigma^{(I)}$ fields we see that the FI couplings
change with the scale and we can define the dynamically generated scales $\Lambda_I$ to be such that 
\begin{equation}
\label{rentauI}
\tau_I(\Lambda_I) = 
\tau_I(\mu)-\frac{k_{I+1}-k_{I-1}}{2\pi\ii}\log\frac{\Lambda_I}{\mu}=0
\end{equation}
for $I=1,\ldots,M-1$\,\footnote{We assume that $k_0=0$ and $k_M=N$.}. 
Equivalently, we can write
\begin{equation}
\Lambda_I^{b_I}=\rme^{2\pi\ii\,\tau_I(\mu)}\,\mu^{b_I}
\label{defLambdaI}
\end{equation}
where 
\begin{equation}
b_I=k_{I+1}-k_{I-1}
\label{bI}
\end{equation}
denotes the coefficient of the $\beta$-function for the running of the FI parameter of the $I$-th 
node.

Using these expressions, the twisted superpotential (\ref{tildeW1}) can thus be rewritten as
\begin{equation}
\begin{aligned}
W=& -\sum_{I=1}^{M-2} \sum_{s=1}^{k_{I}}  \sum_{t=1}^{k_{I+1}}
\sigma^{(I)}_s\bigg( \log\frac{\sigma^{(I)}_s - \sigma^{(I+1)}_t}{\Lambda_I}-1\bigg) \\
&+\sum_{I=2}^{M-1}
  \sum_{s=1}^{k_{I}}\sum_{r=1}^{k_{I-1}}\sigma^{(I)}_s   \bigg(\log
  \frac{\sigma^{(I-1)}_r-\sigma^{(I)}_s}{\Lambda_{I}} -1\bigg)\\
 &-\sum_{s=1}^{k_{M-1}}\left\langle\Tr \left[(\sigma^{(M-1)}_s - 
\Phi)\bigg(\log\frac{\sigma^{(M-1)}_s - \Phi}{\Lambda_{M-1}}-1\bigg)\right]\right\rangle~.
\end{aligned}
\label{WHat} 
\end{equation}
The $I$-th term ($1\leq I\leq M-2)$) in the first line and the $(I+1)$-th term
in the second line of this expression are obtained by integrating 
out the bifundamental fields between the nodes $I$ and $I+1$, while the last line is the result of integrating out the fundamental fields attached to the last gauge node of the quiver. The angular brackets account for 
the four-dimensional dynamics of the SU($N$) theory.
One can easily verify that for $N=M=2$, the expression in (\ref{WHat}) reduces to (\ref{quantumW}). 

\subsubsection*{The twisted chiral ring}
The twisted chiral ring relations are given by 
\begin{equation}
\label{CReq1}
\exp\Big(\frac{\partial{W}}{\partial \sigma^{(I)}_s} \Big) = 1~.
\end{equation}
In order to write the resulting equations in a compact form, we define a characteristic 
gauge polynomial for each of the SU$(k_I)$ node of the quiver
\begin{equation}
Q_{I}(z) = \prod_{s=1}^{k_{I}} (z - \sigma^{(I)}_s)~.
\end{equation}
For $I=1,\ldots,M-2$, the equations are independent of the four-dimensional theory, and read 
\begin{equation}
Q_{{I+1}}(z) = (-1)^{k_{I-1}} \,\Lambda_{I}^{b_I}\,Q_{{I-1}}(z)
\label{boxedone}
\end{equation}
with $z=\sigma^{(I)}_s$ for each $s$, and with the understanding that $Q_0=1$ and $k_0=0$.
Note that the power of $\Lambda_I$, which is determined by the running of the FI coupling, makes 
the equation consistent from a dimensional point of view. For $I=M-1$, the presence of the four-dimensional
SU($N$) gauge theory affects the last two-dimensional node of the quiver, and the corresponding 
chiral ring equation is
\begin{equation}
\exp{\Big\langle \Tr\log\frac{z -\Phi}{\Lambda_{M-1}} \Big\rangle}
 = (-1)^{k_{M-2}}\,\Lambda_{M-1}^{b_{M-1}-N}\,Q_{{M-2}}(z)
\label{CRN0}
\end{equation}
with $z=\sigma^{(M-1)}_s$ for each $s$.
We now use the fact that the resolvent of the four-dimensional SU($N$) theory, which captures all 
information about the chiral correlators, is given by \cite{Cachazo:2002ry}
\begin{equation}
T(z) := \left\langle\Tr \frac{1}{z-\Phi}\right\rangle = \frac{P_N'(z)}{\sqrt{P_N(z)^2-4\Lambda^{2N}}}
\label{resolvent4d}
\end{equation}
where $P_N(z)$ is the characteristic polynomial of degree $N$ encoding the Coulomb vev's of the SU($N$)
theory and $\Lambda$ is its dynamically generated scale.
Since we are primarily interested in the semi-classical solution of the chiral ring equations, we 
exploit the fact that $P_N(z)$ can be written as a perturbation of the classical gauge polynomial in 
the following way:
\begin{equation}
P_N(z) = \prod_{i=1}^N (z-e_i)
\label{PNdef}
\end{equation}
where $e_i$ are the quantum vev's of the pure SU$(N)$ theory given 
by \cite{DHoker:1996pva,Naculich:2002hi}
\begin{equation}
e_i = a_i  - \Lfour^{2N}\frac{\p}{\p a_i}\bigg(\prod_{j\ne i}\frac{1}{a_{ij}^2} \bigg)+O(\Lfour^{4N}) ~.
\label{ekdef}
\end{equation}
Integrating the resolvent (\ref{resolvent4d}) 
with respect to $z$ and exponentiating the resulting expression, one finds
\begin{equation}
\exp\Big\langle \Tr\log\frac{z -\Phi}{\Lambda_{M-1}}\Big \rangle \,=\,\frac{P_N(z) + \sqrt{P_N(z)^2 
- 4\Lambda^{2N}\phantom{\big|}}}{2\Lambda_{M-1}^N}~.
\end{equation}
Using this, we can rewrite the twisted chiral ring relation (\ref{CRN0}) associated to the 
last node of the quiver in the following form: 
\begin{equation}
P_N(z) + \sqrt{P_N(z)^2 - 4\Lambda^{2N}\phantom{\big|}} 
= 2\,(-1)^{k_{M-2}}\,\Lambda_{M-1}^{b_{M-1}}\, Q_{{M-2}}(z)~,
\label{CRN1}
\end{equation}
where $z=\sigma^{(M-1)}_s$. With further simple manipulations, we obtain
\begin{equation}
P_N(z) =(-1)^{k_{M-2}}\,\Lambda_{M-1}^{b_{M-1}}\, Q_{{M-2}}(z) 
+\frac{\Lambda^{2N}}{ (-1)^{k_{M-2}} \,\Lambda_{M-1}^{b_{M-1}}\,Q_{{M-2}}(z) }
\label{boxedtwo}
\end{equation}
for $z=\sigma^{(M-1)}_s$. In the limit $\Lambda\rightarrow 0$ which corresponds to turning off the
four-dimensional dynamics, we obtain the expected twisted chiral ring relation of the last
two-dimensional node of the quiver. 
Equations (\ref{boxedone}) and (\ref{boxedtwo}) are the relevant chiral relations which we are going to
solve order by order in the $\Lambda_I$'s to obtain the weak-coupling expansion of the twisted
chiral superpotential.

\subsubsection*{Solving the chiral ring equations}
Our goal is to provide a systematic procedure to solve the twisted chiral ring
equations we have just derived and to find the effective twisted superpotential of the
2d/4d theory. As illustrated in the case of the SU(2) theory in Section~\ref{SU2}, we shall do so 
by evaluating ${W}$ on the solutions of the twisted chiral ring equations. Each choice
of vacuum therefore corresponds to a different surface operator.

In order to clarify this last point, we first solve the \emph{classical} chiral ring equations, which are obtained by setting $\Lambda_I$ and $\Lambda$ to zero keeping their ratio fixed, {\it{i.e.}} by considering the theory 
at a scale much bigger than $\Lambda_I$ and $\Lambda$. 
Thus, in this limit the right-hand sides
of (\ref{boxedone}) and (\ref{boxedtwo}) vanish. A possible choice \footnote{All other solutions 
are related to this one by permuting the $a$'s.} that accomplishes this is:
\begin{equation}
\begin{aligned}
\label{chosenvacuum}
\sigma^{(1)}_s =&\, a_s\qquad\mbox{for}~s=1, \ldots, k_1~, \\
\sigma^{(2)}_t =&\, a_t \qquad\mbox{for}~t=1, \ldots, k_2 ~,\\ 
&\qquad\vdots\cr
\sigma^{(M-1)}_w =&\, a_w\qquad\!\mbox{for}~w=1, \ldots, k_{M-1}~.
\end{aligned}
\end{equation}
This is equivalent to assuming that the classical expectation value of $\sigma$ for the $I$-th node
is
\begin{equation}
\sigma^{(I)} = \mathrm{diag}\big(a_1,a_2,\ldots,a_{k_I}\big)~.
\label{sigmavac}
\end{equation}
We will see that this choice is the one appropriate to describe a surface defect
that breaks the gauge group SU$(N)$ according to the Levi decomposition (\ref{Levi}).

Let us now turn to the quantum chiral ring equations. Here we make an ansatz for 
$\sigma^{(I)}$ as a power series in the various $\Lambda_I$'s around the chosen classical vacuum. From the explicit expressions (\ref{boxedone}) and (\ref{boxedtwo}) of the chiral ring equations, it is easy to realize that there is a natural set of parameters in terms of which these power series can be written; they are given by
\begin{equation}
q_I = (-1)^{k_{I-1}}\,\Lambda_I^{b_I} 
\label{qIdefn}
\end{equation}
for $I=1,\ldots,M-1$.
If the four-dimensional theory were not dynamical, these $(M-1)$ parameters
would be sufficient; however, from the chiral ring equations (\ref{boxedtwo}) of the last
two-dimensional node of the quiver, we see that another parameter is needed. It is related to the four-dimensional scale $\Lambda$ and hence to the four-dimensional instanton action.
It turns out that this remaining expansion parameter is 
\begin{equation}
q_M = (-1)^N\,\Lambda^{2N}\,\Big(\prod_{I=1}^{M-1}q_I\Big)^{-1}~.
\label{qMdefn}
\end{equation}
Our proposal is to solve the chiral ring equations (\ref{boxedone}) and (\ref{boxedtwo}) as a simultaneous power series in all the $q_I$'s, including $q_M$, which ultimately will be identified with the Nekrasov-like counting parameters in the ramified instanton computations described in Section~\ref{localize4d5d}.

We will explicitly illustrate these ideas in some examples in the next section, but first we would like to show in full generality that the logarithmic derivatives with respect to $\Lambda_I$ are 
directly related to the solution $\sigma^{(I)}_{\star}$ of the twisted chiral ring equations (\ref{boxedone}) and (\ref{boxedtwo}).
The argument is a straightforward generalization of what we have already seen in the
SU(2) case, see (\ref{Wsplit}) and (\ref{dWdL}). On shell, {\it{i.e.}} when $\partial W/\partial\sigma\big|_{\sigma_{\star}}=0$, the twisted superpotential $W_\star\equiv W(\sigma_\star)$ depends 
on $\Lambda_I$ only explicitly. Using the expression of $W$ given in (\ref{WHat}), we find
\begin{equation}
\label{logLider}
\Lambda_I \frac{d W_\star}{d {\Lambda_I}} 
= \Lambda_I \frac{\partial W}{\partial {\Lambda_I}}\bigg|_{\sigma_\star} = 
b_I\, \tr \sigma^{(I)}_\star~,
\end{equation}  
where in the last step we used the tracelessness of $\Phi$. This relation can be written in terms of the parameters $q_I$ defined in (\ref{qIdefn}), as follows
\begin{equation}
\label{logqider}
q_I \frac{d W_\star}{d {q_I}} %= q_I \frac{\partial W_\star}{\partial {q_I}} 
= \tr \sigma^{(I)}_\star~.
\end{equation}  
If we express the solution $\sigma_\star$ of the chiral ring equations as the classical solution 
(\ref{chosenvacuum}) plus quantum corrections, we find
\begin{equation}
\label{Wclcorr}
\begin{aligned}
q_1 \frac{d W_\star}{d {q_1}} & = a_1 + \ldots a_{k_1}+~\mbox{corr.ns} ~
= ~a_1 + \ldots + a_{n_1}
+~\mbox{corr.ns}
~,\\
q_2 \frac{d W_\star}{d {q_2}} & = a_1 + \ldots a_{k_2} +~\mbox{corr.ns} ~
= ~a_1 + \ldots + a_{n_1+n_2} +~\mbox{corr.ns}
\end{aligned}
\end{equation}
and so on.
This corresponds to a partition of the classical vev's of the SU$(N)$ theory given by
\begin{equation}
\left\{
\begin{array}{ccccccccc}
\undermat{n_1}{a_1, & \cdots & a_{n_1},} & \undermat{n_2}{ a_{n_1+1}, & \cdots&a_{n_1+n_2},} &\cdots, 
&\undermat{n_M}{ a_{N-n_M+1}, &\ldots a_{N} }
\end{array}
\right\} ~,
\label{agen}
\vspace{.4cm}
\end{equation}
which is interpreted as a breaking of the gauge group SU$(N)$ according to the Levi decomposition
(\ref{Levi}).
In fact, by comparing with the results of \cite{Ashok:2017odt} (see for instance, equation (4.1) of this reference), we see that the expressions (\ref{Wclcorr}) coincide with the derivatives of the classical superpotential describing the surface operator
of the SU($N$) theory, labeled by the partition $[n_1,n_2,\ldots,n_M]$, provided
we relate the parameters $q_I$ to the variables $t_J$ that label the monodromy defect
according to
\begin{equation}
\label{tItoqI}
2\pi\ii \,t_J \sim  \sum_{I=I}^{M} \log q_I \qquad\text{for}~ J = 1, \ldots, M~.
\end{equation}
We now illustrate these general ideas in a few examples.

\subsection{SU(3)}

We consider the surface operators in the SU(3) theory. There are two 
distinct partitions, namely $[1,2]$ and $[1,1,1]$, which we now discuss in detail.

\subsubsection*{SU(3)[1,2]}
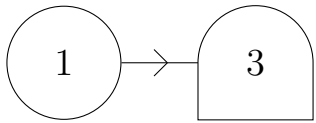
\begin{figure}[ht]
\begin{center}
\begin{tikzpicture}[decoration={markings,
mark=at position 0.6 with {\draw (-5pt,-5pt) -- (0pt,0pt);
                \draw (-5pt,5pt) -- (0pt,0pt);}}]
  \matrix[row sep=10mm,column sep=5mm] {
      \node(g1)[gauge] {\Large $1$};      & &\node(gf2)[gaugedflavor]{\Large $3$};\\
 };
\graph{(g1) --[postaction={decorate}](gf2);};
\end{tikzpicture}
\end{center}
\vspace{-0.5cm}
\caption{The quiver corresponding to the surface operator SU(3)[1,2].}
\label{quiversu312}
\end{figure}

In this case the two-dimensional theory is a U(1) gauge theory with three flavors, represented by the quiver
in Fig.~\ref{quiversu312}.

\noindent
Since $M=2$, we have just one $\sigma$ and one chiral ring equation, which is given by (see (\ref{boxedtwo}))
\begin{equation}
P_3(\sigma) = \Lambda_1^3 + \frac{\Lfour^{6}}{\Lambda_1^3 }
\label{TCR13}
\end{equation}
where the gauge polynomial is defined in (\ref{PNdef}). We solve this equation order by order in
$\Lambda_1$ and $\Lambda$, using the ansatz 
\begin{equation}
\sigma_\star = a_1 + \sum_{\ell_1, \ell_2} c_{\ell_1, \ell_2}\, q_1^{\ell_1}\,q_2^{\ell_2} 
\label{sigmasu312}
\end{equation}
where the expansion parameters are defined in (\ref{qIdefn}) and (\ref{qMdefn}), which in this case
explicitly read
\begin{equation}
q_1 = \Lambda_1^3~,\qquad q_2= -\frac{\Lfour^6}{\Lambda_1^3}~.
\label{qISU3}
\end{equation}
Inserting (\ref{sigmasu312}) into (\ref{TCR13}), we can recursively determine the coefficients
$c_{\ell_1\ell_2}$ and, at the first orders, find the following result
\begin{equation}
\sigma_\star=a_1+ 
\frac{1}{a_{12}\,a_{13}}\Big(\Lambda _1^3 + 
\frac{\Lfour^6}{\Lambda_1^3}\Big)-\Big(\frac{1}{a_{12}^3\,a_{13}^2}
+\frac{1}{a_{12}^2\,a_{13}^3}\Big)\Big(\Lambda _1^6 + 
\frac{\Lfour^{12}}{\Lambda_1^6}\Big) +\ldots
\label{4d[1,3]answer}
\end{equation}
where $a_{ij}=a_i-a_j$. According to (\ref{logqider}),
this solution coincides with the 
$q_1$-logarithmic derivative
of the twisted superpotential. We will verify this statement by comparing (\ref{4d[1,3]answer}) against 
the result obtained via localization methods.

\subsubsection*{SU(3)[1,1,1]}
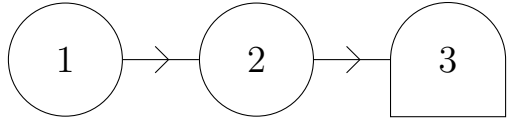
\begin{figure}[ht]
\begin{center}
\begin{tikzpicture}[decoration={markings,
 mark=at position 0.6 with {\draw (-5pt,-5pt) -- (0pt,0pt);
                \draw (-5pt,5pt) -- (0pt,0pt);}}]
  \matrix[row sep=10mm,column sep=5mm] {
      \node(g1)[gauge] {\Large $1$};  & & \node(g2)[gauge] {\Large $2$}; 
      & &\node(gfN)[gaugedflavor]{\Large $3$};\\
 };
\graph{(g1) --[postaction={decorate}](g2)--[postaction={decorate}](gfN);};
\label{quiverpic112}
\end{tikzpicture}
\end{center}
\vspace{-0.5cm}
\caption{The quiver diagram representing the surface operator SU(3)[1,1,1].}
\label{quiversu3111}
\end{figure}

In this case the two-dimensional theory is represented by the quiver of Fig.~\ref{quiversu3111}.

\noindent
Since $M=3$, there are now two sets of twisted chiral ring equations. For the first node, from
(\ref{boxedone}) we find
\begin{equation}
\prod_{s=1}^2\big(\sigma^{(1)} - \sigma^{(2)}_s\big) = \Lambda_1^2~,
\label{CR1}
\end{equation}
while for the second node, from (\ref{boxedtwo}) we get
\begin{equation}
P_3\big(\sigma_s^{(2)}\big) = -\Lambda_2^2 \,\big(\sigma_s^{(2)} - \sigma^{(1)}\big) 
- \frac{\Lfour^6}{\Lambda_2^2 \,\big(\sigma_s^{(2)}  - \sigma^{(1)}\big)}
\label{CR2}
\end{equation}
for $s=1,2$.
{From} the classical solution to these equations (see (\ref{sigmavac})), we realize that this configuration
corresponds to a surface operator specified by the partition of the Coulomb 
vev's $\{\{a_1\},\{a_2\},\{a_3\}\}$, which is indeed associated to the partition [1,1,1] we are considering. 
Thus, the ansatz for solving
the quantum equations (\ref{CR1}) and (\ref{CR2}) takes the following form:
\begin{equation}
\begin{aligned}
\sigma^{(1)}_\star &= a_1 + \sum_{\ell_1, \ell_2,\ell_3} d_{\ell_1, \ell_2, \ell_3} \,\,q_1^{\ell_1}
\,q_2^{\ell_2}\,q_3^{\ell_3}~,\\
\sigma^{(2)}_{\star,1} &= a_1 + \sum_{\ell_1, \ell_2,\ell_3} f_{\ell_1, \ell_2, \ell_3} \,\,q_1^{\ell_1}
\,q_2^{\ell_2}\,q_3^{\ell_3}~,\\
\sigma^{(2)}_{\star,2} &= a_2 + \sum_{\ell_1, \ell_2,\ell_3} g_{\ell_1, \ell_2, \ell_3} \,\,q_1^{\ell_1}
\,q_2^{\ell_2}\,q_3^{\ell_3}~,
\end{aligned}
\end{equation}
with
\begin{equation}
q_1 = \Lambda_1^2~,\quad
q_2 =- \Lambda_2^2~,\quad
q_3 = \frac{\Lfour^6}{\Lambda_1^2\Lambda_2^2}~.
\label{q123}
\end{equation}
Solving the coupled equations (\ref{CR1}) and (\ref{CR2}) order by order in $q_I$, 
we find the following result:
\begin{align}
\sigma^{(1)}_\star=& \,\,a_1+ \frac{1}{a_{12}}\,\Lambda _1^2
+\frac{1}{a_{13}}\,\frac{\Lambda^6}{\Lambda _1^2\Lambda_2^2}
-\frac{1}{a_{12}^3}\,\Lambda _1^4
-\frac{1}{a_{13}^3}\,\frac{\Lambda ^{12}}{\,\Lambda _1^4 \Lambda _2^4}\notag
\\
&\qquad\quad
-\frac{1}{a_{12}\,a_{13}\,a_{23}}\Big(\Lambda _1^2\Lambda _2^2-\frac{\Lambda ^6}{
\Lambda _1^2}\Big)+\ldots ~,
\label{sigma1}\\
\notag\\
\Tr \sigma^{(2)}_{\star}=& \,\,a_1+a_2-\frac{1}{a_{23}}\,\Lambda _2^2
+\frac{1}{a_{13}}\,\frac{\Lambda ^6}{\Lambda _1^2\Lambda _2^2}
-\frac{1}{a_{23}^3}\,\Lambda _2^4
-\frac{1}{a_{13}^3}\,\frac{\Lambda ^{12}}{\Lambda_1^4\Lambda _2^4 }\notag\\
&\qquad\quad
-\frac{1}{a_{12}\,a_{13}\,a_{23}}\Big(\Lambda _1^2\Lambda _2^2 
+\frac{\Lambda ^6}{\,\Lambda _2^2 }\Big)+\ldots~.
\label{sigma12}
\end{align}
According to (\ref{logqider}) these expressions should be identified, respectively, with the 
$q_1$- and
$q_2$-logarithmic derivatives of the twisted superpotential. We will verify this relation in 
Section~\ref{localize4d5d}
using localization.

We have analyzed in detail the SU(3) theory in order to exhibit how explicit and systematic our methods
are. We have thoroughly explored all surface defects in the SU(4) and SU(5) theories and also 
considered a few other examples with higher rank gauge groups. In all these cases our method 
of solving the twisted chiral ring equations proved to be very efficient and quickly led to very explicit results.
One important feature of our approach is the choice of classical 
extrema of the twisted superpotential which will allow us to make direct contact with the localization 
calculations of the superpotential for Gukov-Witten defects in four-dimensional gauge theories. A further essential ingredient is the use of the 
quantum corrected resolvent in four dimensions, which plays a crucial role in obtaining the 
higher-order solutions of the twisted chiral ring equations of the two-dimensional quiver theory.

\section{Twisted superpotential for coupled 3d/5d theories}
\label{quiver3d5d}
Let us now consider the situation in which the 2d/4d theories described in the previous section 
are replaced by 3d/5d ones compactified on a circle $S^1_\beta$ of length $\beta$. The content of 
these theories is still described by quivers of the same form as in Fig.~\ref{quiverpicgeneric1}. 
We begin by considering the three-dimensional part. 

\subsection{Twisted chiral ring in quiver gauge theories}
To construct the effective theory for the massless chiral twisted fields, which is encoded in the twisted 
superpotential $W$, we have to include the contributions of all Kaluza-Klein (KK) copies of 
the (bi)-fundamental matter multiplets. When the scalars $\sigma^{(I)}$ are gauge-fixed as in (\ref{sigmadiag}), 
the KK copies of the matter multiplets have masses\,\footnote{\label{foot:shiftsigma}This is consistent 
with the fact that gauge-fixing the scalars $\sigma^{(I)}$ as in (\ref{sigmadiag}) leaves 
a residual invariance under which the eigenvalues shift by $\sigma^{(I)}_s \to \sigma^{(I)}_s 
+ {2\pi\ii n_s}/{\beta}$.} 
\begin{equation}
\label{KKcopiesmm}
\sigma^{(I)}_s - \sigma^{(I+1)}_t + {2\pi\ii \,n}/{\beta}~,
\end{equation}
for $I=1,\ldots,M-2$ (with an independent integer $n$ for each multiplet).
Similarly, the copies of the matter multiplet attached to the 5d node have masses
\begin{equation}
\sigma^{(M-1)}_s - a_i + {2\pi\ii \,n}/{\beta}
\end{equation}
when the 5d theory is treated classically.   

All these chiral massive fields contribute to the one-loop part of $W$. As we saw in (\ref{tildeW1})
and (\ref{defvarpi}), in 2d a chiral field of mass $z$ contributes a term proportional to $\varpi(z)$.  
Summing over all its KK copies results therefore in a contribution proportional to
\begin{equation}
\label{defell}
\ell(z) \equiv \sum_{n\in\mathbb{Z}} \varpi\big(z + 2\pi\ii \,n/\beta\big)
\end{equation}
where the sum has to be suitably regularized.
This function satisfies the property
\begin{equation}
\label{derell}
\partial_z\ell(z) = \sum_{n\in\mathbb{Z}} 
\partial_z\varpi\big(z+ 2\pi\ii \,n/\beta\big) =  \sum_{n\in\mathbb{Z}} 
\log \Big(\frac{z + 2\pi\ii \,n/\beta}{\mu}\Big)
= \log\Big(2 \sinh \frac{\beta z}{2}\Big)~.
\end{equation}
Note that the scale $\mu$, present in the definition (\ref{defvarpi}) of the function $\varpi(z)$, no 
longer appears after the sum over the KK copies.  Integrating this relation, one gets
\begin{equation}
\label{ellis}
\ell(z) = \frac{1}{\beta}\, \mathrm{Li}_2\big(\rme^{-\beta z}\big) + 
\frac{\beta z^2}{4} - \frac{\pi^2}{6\beta}~,
\end{equation}
where the integration constant has been fixed in such a way that 
\begin{equation}
\label{limell}
\ell(z) ~\stackrel{\beta\to 0}{\sim}~ z\,\Big(\log (\beta z) - 1\Big) = \varpi(z)~.
\end{equation}
Note that here $\varpi(z)$ is defined taking the UV scale to be 
\begin{equation}
\label{muisbeta}
\mu = 1/\beta~,
\end{equation} 
as is natural in this compactified situation.

Therefore, the twisted superpotential of the three-dimensional theory is simply given by (\ref{tildeW1}) 
with all occurrences of the function $\varpi(z)$ replaced by $\ell(z)$, for any argument $z$, and with 
the UV scale $\mu$ being set to $1/\beta$. Just as in the two-dimensional 
case, we would like to replace the FI couplings at the UV scale, $\tau_I(1/\beta)$, 
with the dynamically generated scales $\Lambda_I$. 
Since the renormalization of these FI couplings is determined only by the lightest KK multiplets, the running is the same as in two dimensions and thus we can simply use (\ref{rentauI}) with $\mu$ 
identified with $1/\beta$ according to (\ref{muisbeta}).
Note however that, in contrast to the two-dimensional case described in
(\ref{WHat}), this replacement does not eliminate completely the UV scale from the expression of $W$, and 
the dependence on $\beta$ remains in the functions $\ell(z)$. Altogether we have 
\begin{equation}
\begin{aligned}
{W}=&\sum_{I=1}^{M-1}\sum_{s=1}^{k_I} b_I \log (\beta\Lambda_I)
\sigma^{(I)}_s
- \!\sum_{I=1}^{M-2} \sum_{s=1}^{k_{I}} \sum_{t=1}^{k_{I+1}}
\ell\big(\sigma^{(I)}_s - \sigma^{(I+1)}_t\big)
-\!\!\sum_{s=1}^{k_{M-1}}\!\Big\langle 
\Tr \ell\big(\sigma^{(M-1)}_s - \Phi\big) \Big\rangle~.
\end{aligned}
\label{tildeW15d}
\end{equation}
The expectation value in the last term is taken with respect to the five-dimensional 
gauge theory defined on the last node of the quiver and compactified on the same circle of length 
$\beta$ as the three-dimensional sector. 

\subsubsection*{The twisted chiral ring}

Our aim is to show that the twisted superpotential (\ref{tildeW15d}), evaluated on a suitably 
chosen vacuum $\sigma^\star$, matches the twisted superpotential extracted via localization 
for a corresponding monodromy defect. 
Just as in the 2d/4d case, the vacuum $\sigma^\star$ minimizes $W$, namely solves 
the twisted chiral ring equation (\ref{CReq1}). Moreover, the logarithmic derivatives of 
$W$ with respect to $\Lambda_I$, or with respect to the parameters $q_I$ in
(\ref{qIdefn}), evaluated on a solution $\sigma^\star$, still satisfy (\ref{logLider}) or
(\ref{logqider}) respectively. These derivatives are quite simple to compute and these are the 
quantities that we will compare with localization results.  

In close parallel to what we did in the 2d/4d case, the chiral ring equations (\ref{CReq1}) can be expressed in a compact form if we introduce the quantity
\begin{equation}
\widehat{Q}_{I}(z) = \prod_{s=1}^{k_{I}}\Big( 2 \sinh\frac{\beta(z - \sigma^{({I})}_{s})}{2}\Big)
\label{tildeQI}
\end{equation}
for each of the SU$(k_I)$ gauge groups in the quiver; note that $\widehat{Q}_{I}(z)$ is naturally written 
in terms of the exponential variables
\begin{equation}
\label{defS}
S^{(I)}_s = \rme^{\beta\sigma^{(I)}_s}~,
\end{equation}
which are invariant under the shifts described in footnote~\ref{foot:shiftsigma}.
Indeed, starting from (\ref{CReq1}) and taking into account (\ref{derell}), 
for $I=1,\ldots,M-2$ we find
\begin{equation}
\widehat{Q}_{{I+1}}(z) = (-1)^{k_{I-1}} \,\big(\beta\Lambda_{I}\big)^{b_I}\,
\widehat{Q}_{{I-1}}(z)
\label{CR3d}
\end{equation}
with $z=\sigma^{(I)}_s$. For the node $I=M-1$ we obtain
\begin{equation}
\exp{\Big\langle \Tr\log\Big(2\sinh\frac{\beta(z -\Phi)}{2}\Big)\Big\rangle}
 = (-1)^{k_{M-2}}\,\big(\beta\Lambda_{M-1}\big)^{b_{M-1}}\,\widehat{Q}_{{M-2}}(z)
\label{CR5d}
\end{equation}
with $z=\sigma^{(M-1)}_s$.
To proceed further, we need to evaluate in the compactified 5d theory the expectation value appearing in 
the left hand side of (\ref{CR5d}). To do so, let us briefly recall a few facts about this compactified gauge theory.

\subsubsection*{The resolvent in the compactified 5d gauge theory}
The five-dimensional ${\mathcal N}=1$ vector multiplet consists of a gauge field $A_{\mu}$, 
a real scalar $\phi$ and a gluino $\lambda$. Upon circle compactification, the component 
$A_t$ of the gauge field along the circle and the scalar $\phi$ give rise to the complex 
adjoint scalar $\Phi = A_t + \ii \phi$ of the four-dimensional $\mathcal{N}=2$ theory. The Coulomb
branch of this theory is classically specified by fixing the gauge \cite{Nekrasov:1996cz}:
\begin{equation}
\label{5dscalar}
\Phi= A_t + \ii \phi = \diag{(a_1, a_2, \ldots a_N)}~.
\end{equation}
However, there is a residual gauge symmetry under which
\begin{equation}
a_i \rightarrow a_i + {2\pi \ii\, n_i}/{\beta}
\label{shift}
\end{equation}
with $n_i\in\mathbb{Z}$; since we are considering a $\mathrm{SU}(N)$ theory, we must ensure that
these shifts preserve the vanishing of $\sum_i a_i$.

The low-energy effective action can be determined in terms of an algebraic curve and a differential, just 
as in the usual four-dimensional case. The Seiberg-Witten curve for this model was first proposed 
in \cite{Nekrasov:1996cz} and later derived from a saddle point analysis of the instanton partition function in 
\cite{Nekrasov:2002qd,Nekrasov:2003rj}; it takes the following form
\begin{equation}
\label{SW5d}
y^2 = \widehat{P}_N^2(z) - 4\big(\beta\Lfour\big)^{2N} ~.
\end{equation}
Here $\Lambda$ is the strong-coupling scale that is dynamically generated and
\begin{equation}
\widehat{P}_N(z) = \prod_{i=1}^{N}\Big( 2 \sinh\frac{\beta(z - e_i)}{2}\Big)
\label{P5ddefn}
\end{equation}
where $e_i$ parametrize the quantum moduli space and reduce to $a_i$ in the classical regime, 
in analogy to the four-dimensional case. Like the latter, they also satisfy a tracelessness condition:
$\sum_i e_i = 0$. Note that $\widehat{P}_N$ can be written purely in terms of the exponential variables
\begin{equation}
\label{expvar}
E_i = \rme^{\beta e_i}~,~~~ Z = \rme^{\beta z}~,
\end{equation}
and is thus invariant under the shift (\ref{shift}). Indeed, using (\ref{expvar}) we find
\begin{equation}
\label{P5ddefexp}
\widehat{P}_N(z) = Z^{-\frac{N}{2}}\Big( Z^N + \sum_{i=1}^{N-1} (-1)^i Z^{N-i}\, U_i + (-1)^N \Big) ~,
\end{equation}
where $U_i$ is the symmetric polynomial
\begin{equation}
U_i = \sum_{j_1< j_2< \ldots j_i} E_{j_1} \ldots E_{j_k}~.
\label{Ukdefn}
\end{equation}
In (\ref{P5ddefexp}) we have used the SU($N$) tracelessness condition, which implies 
$U_N = \prod_i E_i = 1$. 

The resolvent of this five-dimensional theory, defined as \cite{Wijnholt:2004rg}
\begin{equation}
\widehat{T}(z) = \Big\langle\Tr\coth\frac{\beta(z-\Phi)}{2}\Big\rangle
=\frac{2}{\beta} \,\frac{\partial}{\partial z} \Big\langle
\Tr\log \Big(2\sinh\frac{\beta (z-\Phi)}{2}\Big)\Big\rangle~,
\label{resolventdefn}
\end{equation}
contains the information about the chiral correlators through the expansion
\begin{equation}
\widehat{T}(z) = N + 2 \sum_{\ell=1}^{\infty}\rme^{-\ell\beta z}\,\Big\langle
\Tr e^{\ell\, \beta\Phi}\Big\rangle~.
\label{Texp}
\end{equation}
On the other hand, the Seiberg-Witten theory expresses this resolvent 
as\,\footnote{\label{foot:Utochir}Using (\ref{P5ddefexp}) 
we can expand this expression in inverse powers of $Z$; then, comparing to (\ref{Texp}), we can relate 
the correlators $\vev{\Tr e^{\ell\, \beta\Phi}}$, of which the first $(N-1)$ ones are independent, to 
the $U_\ell$'s.}
\begin{equation}
\widehat{T}(z) = \frac{2}{\beta}\,\frac{\widehat{P}_N^\prime(z)}{\sqrt{\widehat{P}_N^2(z) 
- 4\left(\beta \Lfour\right)^{2N}}}~,
\label{5dresolvent}
\end{equation}
so that, integrating (\ref{resolventdefn}), we have
\begin{equation}
\begin{aligned}
\exp\Big\langle\Tr\log \Big(2\sinh\frac{\beta (z-\Phi)}{2}\Big)\Big\rangle 
& = \frac{\widehat{P}_N(z) + \sqrt{\widehat{P}_N^2(z)- 4(\beta\Lfour)^{2N}}}{2 } ~.
\end{aligned}
\label{resolventintegral}
\end{equation}
With manipulations very similar to those described in Section~\ref{quiver2d4d} for the 2d/4d case, 
we can now rewrite the twisted chiral ring relation (\ref{CR5d}) as follows
\begin{equation}
\begin{aligned}
\widehat{P}_N(z) &=(-1)^{k_{M-2}}\,\big(\beta\Lambda_{M-1}\big)^{b_{M-1}}\, 
\widehat{Q}_{{M-2}}(z) \\
&\qquad\qquad\qquad\qquad\qquad+\frac{\big(\beta\Lambda\big)^{2N}}{ (-1)^{k_{M-2}} 
\,\big(\beta\Lambda_{M-1}\big)^{b_{M-1}}\,
\widehat{Q}_{{M-2}}(z) }
\end{aligned}
\label{CR5dfin}
\end{equation}
for $z=\sigma^{(M-1)}_s$.  It is easy to check that in the limit $\beta\rightarrow 0$ we recover 
the corresponding equation (\ref{boxedtwo}) for the 2d/4d theory.

\subsection*{Solving the chiral ring equations}
At the classical level the solution to the chiral ring equations
takes exactly the same form as in (\ref{sigmavac}). In terms of the exponential variables 
introduced in (\ref{defS}) we can write it as
\begin{equation}
\label{sigmavace}
S^{(I)}_\star= \mathrm{diag}(A_1,\ldots,A_{k_I})
\end{equation}
where $A_i=\rme^{\beta a_i}$. These variables $A_i$ 
represent the classical limit of the variables $E_i$ defined in (\ref{expvar}). The SU($N$) tracelessness
condition implies that $\prod_iA_i=1$.

Our aim is to solve the chiral ring equations (\ref{CR3d}) and (\ref{CR5dfin}), and then compare 
the solutions to the localizations results, which naturally arise in a semi-classical expansion. 
Therefore, we propose 
an ansatz that takes the form of an expansion in powers of $\beta$, namely
\begin{equation}
\label{ansS}
S^{(I)}_\star = \mathrm{diag}\Big(A_1+\sum_\ell \delta S^{(I)}_{1,\ell},\ldots,
A_{k_I}+\sum_\ell \delta S^{(I)}_{k_I,\ell}\Big) ~.
\end{equation}
Notice that also the chiral ring equation (\ref{CR5dfin}) of the last node can be expanded in $\beta$. Indeed,
the quantity $\widehat{P}_N$ contains the moduli space coordinates $U_i$, which as shown in 
Appendix~\ref{app:chiralcorr5d}, admit a natural expansion in powers of $\big(\beta\Lambda\big)^{2N}$.
Putting everything together, we can solve all chiral ring equations iteratively, order by order
in $\beta$ and determine the corrections $\delta S^{(I)}_{s,\ell}$ and thus the solution $S^{(I)}_\star$. In this
way, repeating the same steps of the 2d/4d theories, we obtain the expression of the
logarithmic derivative of the twisted superpotential, namely
\begin{equation}
q_I\frac{d {W}_\star}{d q_I}
= \frac{1}{\beta}\sum_{s=1}^{k_I} \log S^{(I)}_{\star,s}=\Tr \sigma^{(I)}_{\star}~.
\label{derWsol}
\end{equation}

\subsection{SU(2) and SU(3)}
We now show how this procedure works in a few simple examples with gauge groups of low rank.

\subsubsection*{SU(2)[1,1]} 
In this case the quiver is the one drawn in Fig.~\ref{quiversu2}. Since $M=2$, there 
is a single variable $\sigma$ for the U$(1)$ node and a single FI parameter $\tau$. 
The only chiral ring equation is given by
(\ref{CR5dfin}) with $z=\sigma$, namely
\begin{equation}
\label{CR5dsu2}
\widehat{P}_2(\sigma) = \beta^2 \Big(\Lambda_1^2 + \frac{ \Lfour^4}{\Lambda_1^2}\Big) ~.
\end{equation} 
Using (\ref{P5ddefexp}) we can express $\widehat{P}_2$ in terms of $S=\rme^{\beta\sigma}$, obtaining
\begin{equation}
\label{P2is}
\widehat{P}_2(S) = S + \frac{1}{S} - U_1 
= 2 \cosh(\beta\sigma) - U_1
\end{equation}
where $U_1=E_1+E_2$. A solution of the twisted chiral ring equation is therefore given by
\begin{equation}
\label{solSsu2}
\sigma_\star = \frac{1}{\beta}\,\log S_\star = 
\frac{1}{\beta} \,\mathrm{arccosh}
\left[\frac{U_1}{2}+  \frac{\beta^2}{2}
 \Big(\Lambda_1^2 + \frac{ \Lfour^4}{\Lambda_1^2}\Big)\right]~.
\end{equation}
In Appendix~\ref{app:chiralcorr5d} we derive the semi-classical expansion of $U_1$. This is 
given in (\ref{U12}) which, rewritten in terms of $a$, reads
\begin{equation}
U_1 =2\cosh(\beta a)\, \bigg(1+ \frac{\big(\beta\Lambda\big)^4}{4\sinh^2(\beta a)}+ \ldots \bigg)~.
\end{equation}
Substituting this into (\ref{solSsu2}), we find finally
\begin{equation}
\label{sigmasol}
\sigma_\star = a + \frac{\beta}{2\sinh(\beta a)}\Big(\Lambda_1^2+\frac{\Lfour^4}{\Lambda_1^2}
\Big)-\frac{\beta^3\cosh(\beta a)}{8\sinh^3(\beta a)}
\Big(\Lambda_1^4+\frac{\Lfour^8}{\Lambda_1^4}\Big)+\ldots~.
\end{equation}
According to (\ref{derWsol}), this solution corresponds to the logarithmic $q_1$-derivative of the superpotential, namely
\begin{equation}
\label{dwissigma}
q_1\frac{d  W_\star}{d q_1} = \sigma_\star~.
\end{equation}
We will verify in the next section that this is indeed the case, by comparing with the superpotential computed
via localization and finding a perfect match.

\subsubsection*{SU(3)[1,2]}
This case is described by the quiver in Fig.~\ref{quiversu312}. Again, we have $M=2$ and thus
a single variable $\sigma$ and a single FI parameter $\tau$. In this case, the chiral ring 
equation (\ref{CR5dfin}) reads 
\begin{equation}
\label{CR5dsu3}
\widehat{P}_3(S) = \beta^3 \Big(\Lambda_1^3 + \frac{ \Lfour^6}{\Lambda_1^3}\Big)
\end{equation} 
where
\begin{equation}
\label{P35d}
\widehat{P}_3(S) = S^{-{3}/{2}}\big(S^3-U_1 S^2 + U_2 S - 1 \big)~.
\end{equation}
Using the semi-classical expansions of $U_1$ and $U_2$ given in (\ref{U13}) and (\ref{U23}), 
and solving the chiral ring equation order by order in $\beta$ according to the ansatz (\ref{ansS}), 
we obtain 
\begin{equation}
\begin{aligned}
\label{dW135d}
\sigma_\star =\frac{1}{\beta}\log S_\star =&\, a_1 + \beta^2 \frac{A_1^{{1}/{2}}}{A_{12}A_{13}}
\Big(\Lambda_1^3+\frac{\Lambda^6}{\Lambda_1^3} \Big)\\
&~-\frac{\beta^5}{2}\Big(\frac{A_1(A_1+A_2)}{A_{12}^3A_{13}^2}+\frac{A_1(A_1+A_3)}{A_{12}^2A_{13}^3}\Big) 
\Big(\Lambda_1^6+\frac{\Lambda^{12}}{\Lambda_1^6} \Big)+\ldots
\end{aligned}
\end{equation}
where $A_{ij}=A_i-A_j$. Rewriting this solution in terms of the classical vev's $a_i$, we have
\begin{align}
\sigma_\star =&\, a_1 +  \frac{\beta^2}{4 \sinh\big(\frac{\beta}{2} a_{12}\big)
\sinh\big(\frac{\beta}{2} a_{13}\big)}
\Big(\Lambda_1^3+\frac{\Lambda^6}{\Lambda_1^3} \Big)\label{dW135da}\\
&~-\frac{\beta^5}{32}\Big(\frac{\cosh\big(\frac{\beta}{2} a_{12}\big)}{\sinh^3\big(\frac{\beta}{2} 
a_{12}\big)\sinh^2\big(\frac{\beta}{2} a_{13}\big)}
\!+\!\frac{\cosh\big(\frac{\beta}{2} a_{13}\big)}{\sinh^2\big(\frac{\beta}{2} 
a_{12}\big)\sinh^3\big(\frac{\beta}{2} a_{13}\big)}\Big) 
\Big(\Lambda_1^6+\frac{\Lambda^{12}}{\Lambda_1^6} \Big)+\ldots\notag
\end{align}
It is very easy to see that in the limit $\beta\to 0$ this reduces to the solution of the corresponding 
2d/4d theory given in (\ref{4d[1,3]answer}). In the next section we will recover this same result 
by computing the $q_1$-logarithmic derivative of the twisted superpotential using localization.

\subsubsection*{SU(3)[1,1,1]}
In this case the quiver is the one drawn in Fig.~\ref{quiversu3111}. Since $M=3$, we have two FI parameters
and two sets of chiral ring equations. For the first node the equation is given by (\ref{CR3d}) which,
in terms of the exponential variables, explicitly reads
\begin{equation}
\prod_{s=1}^2 (S^{(1)} - S^{(2)}_s) = \beta^2\, S^{(1)} \sqrt{S^{(2)}_1 S^{(2)}_2}
\, \Lambda_1^2~.
\end{equation}
For the last node, instead, the chiral ring equations are given by (\ref{CR5dfin}), namely
\begin{equation}
\widehat{P}_3\big(S^{(2)}_s\big) = -\beta^2\left( 
\Lambda_2^2\,\frac{S^{(2)}_s-S^{(1)}}{\sqrt{S^{(1)}S^{(2)}_s}}  +  
\frac{\beta^2\Lambda^6}{\Lambda_2^2}\,
\frac{\sqrt{S^{(1)}S^{(2)}_s}}{S^{(2)}_s-S^{(1)}}
\right)
\end{equation}
for $s=1,2$. Here $\widehat{P}_3$ is as in (\ref{P35d}) with $U_1$ and $U_2$ given 
in (\ref{U13}) and (\ref{U23}). Solving these equations by means of the ansatz (\ref{ansS}), we obtain
\begin{equation}
\sigma^{(1)}_\star = \frac{1}{\beta}\log S^{(1)}_\star=a_1+\beta\,
\frac{\sqrt{A_1 A_2}}{A_{12}}\Lambda_1^2+\beta
\frac{\sqrt{A_1A_3}}{A_{13}}\,\frac{\Lambda^6}{\Lambda_1^2\Lambda_2^2}
+\ldots~,
\end{equation}
and
\begin{equation}
\begin{aligned}
\Tr \sigma^{(2)}_{\star}&=\frac{1}{\beta}\Big(\log S^{(2)}_{\star,1}
+\log S^{(2)}_{\star,2}\Big)\\
&=a_1+a_2-\beta\,
\frac{\sqrt{A_2 A_3}}{A_{23}}\Lambda_2^2+\beta\,\frac{\sqrt{A_1 A_3}}{A_{13}}
\,\frac{\Lambda^6}{\Lambda_1^2\Lambda_2^2}+\ldots~.
\end{aligned}
\end{equation}
 In terms of the classical vev's $a_i$ these solutions become, respectively,
 \begin{equation}
\sigma^{(1)}_\star=a_1+ \frac{\beta}{2\sinh\big(\frac{\beta}{2}a_{12}\big)}\Lambda_1^2+
\frac{\beta}{2\sinh\big(\frac{\beta}{2}a_{13}\big)}\,\frac{\Lambda^6}{\Lambda_1^2\Lambda_2^2}
+\ldots~,
\label{sol1su3111}
 \end{equation}
and
\begin{equation}
\Tr \sigma^{(2)}_{\star}=a_1+a_2-
 \frac{\beta}{2\sinh\big(\frac{\beta}{2}a_{23}\big)}\Lambda_2^2+
\frac{\beta}{2\sinh\big(\frac{\beta}{2}a_{13}\big)}\,\frac{\Lambda^6}{\Lambda_1^2\Lambda_2^2}
+\ldots~.
\label{sol2su3111}
\end{equation}
In the limit $\beta\to 0$ these expressions reproduce the first few terms of the 2d/4d solutions
(\ref{sigma1}) and (\ref{sigma12}) and, as we will see in the next section, they perfectly agree
with the $q_I$-logarithmic derivatives of the twisted superpotential calculated using localization, confirming
(\ref{derWsol}).

We have also computed and checked higher order terms in these SU(3) examples, as well as in theories
with gauge groups of higher rank (up to SU(6)).

\section{Ramified instantons in 4d and 5d}
\label{localize4d5d}

In this section we treat the surface operators as monodromy defects $D$. We begin by considering
the four-dimensional case and later we will discuss the extension to a five-dimensional theory compactified on
a circle of length $\beta$. 

\subsection{Localization in 4d}
We parametrize $\mathbb{R}^4\simeq \mathbb{C}^2$ by two complex variables
$(z_1,z_2)$ and place $D$ at $z_2=0$, filling the $z_1$ plane. The presence of the surface 
operator induces a singular behavior in the gauge connection $A$, which acquires the following
generic form \cite{Alday:2010vg,Kanno:2011fw}:
\begin{equation}
\label{Asing}
A=A_{\mu}\, dx^{\mu}\,\simeq\,-\,\text{diag}
\left(
\begin{array}{cccccccc}
\undermat{n_1}{\gamma_1,\cdots,\gamma_1},\undermat{n_2}{\gamma_2,\cdots,\gamma_2},\cdots, 
\undermat{n_M}{\gamma_M,\cdots,\gamma_M}  
\end{array}
\right)\,d\theta
\vspace{.5cm}
\end{equation}
as $r\to0$. Here $(r,\theta)$ denote the polar coordinates in the $z_2$-plane orthogonal to $D$, and 
$\gamma_I$ are constant parameters that label the surface operator. 
The $M$ integers $n_I$ are a partition of $N$
and identify a vector $\vec{n}$ 
associated to the symmetry breaking pattern of the Levi decomposition (\ref{Levi}) of SU($N$). This vector
also determines the split of the vev's $a_i$ according to (\ref{agen}).

A detailed derivation of the localization results for a generic surface operator has 
been given in \cite{Kanno:2011fw, Nawata:2014nca, Ashok:2017odt},
following earlier mathematical work in \cite{mehta1980, biswas1997, Feigin:2008}. 
Here, we follow the discussion in \cite{Ashok:2017odt} to which we refer for details, 
and present merely those results that are 
relevant for the pure gauge theory. The instanton partition function for a surface operator described 
by $\vec{n}$ is given by\,\footnote{Here, differently from \cite{Ashok:2017odt}, we have introduced
a minus sign in front of $q_I$ in order to be consistent with the conventions chosen in the twisted
chiral ring.}
\begin{equation}
Z_{\text{inst}}[\vec{n}] = \sum_{\{d_I\}}Z_{\{d_I\}}[\vec{n}]\quad
\mbox{with}~~~Z_{\{d_I\}}[\vec{n}]= \prod_{I=1}^M \Big[\frac{(-q_I)^{d_I}}{d_I!}
\int \prod_{\sigma=1}^{d_I} \frac{d\chi_{I,\sigma}}{2\pi\ii}\Big]~
z_{\{d_I\}}
\label{Zso4d5d}
\end{equation}
where
\begin{eqnarray}
z_{\{d_I\}} & =& \,\prod_{I=1}^M \prod_{\sigma,\tau=1}^{d_I}\,
\frac{g\left(\chi_{I,\sigma} - \chi_{I,\tau} + \delta_{\sigma,\tau}\right)
}{g\left(\chi_{I,\sigma} - \chi_{I,\tau} + \epsilon_1\right)}
\times\prod_{I=1}^M \prod_{\sigma=1}^{d_I}\prod_{\rho=1}^{d_{I+1}}\,
\frac{g\left(\chi_{I,\sigma} - \chi_{I+1,\rho} + \epsilon_1 + \hat\epsilon_2\right)}
{g\left(\chi_{I,\sigma} - \chi_{I+1,\rho} + \hat\epsilon_2\right)}
\label{zexplicit5d}
\\
&& ~~\times
\prod_{I=1}^M \prod_{\sigma=1}^{d_I} \prod_{s=1}^{n_I}
\frac{1}
{g\left(a_{I,s}-\chi_{I,\sigma} + \frac 12 (\epsilon_1 + \hat\epsilon_2)\right)}
\prod_{t=1}^{n_{I+1}}
\frac{1}
{g\left(\chi_{I,\sigma} - a_{I+1,t} + \frac 12 (\epsilon_1 + \hat\epsilon_2)\right)}~.
\nonumber
\end{eqnarray}
Let us now explain the notation. The $M$ positive integers $d_I$ count the numbers of ramified
instantons in the various sectors, with the convention that $d_{M+1}=d_1$\,\footnote{Also in $n_I$,
$\chi_I$ and $a_I$, the index $I$ is taken modulo $M$.}. When these numbers 
are all zero, we understand that $Z_{\{d_I=0\}}=1$. The $M$ variables $q_I$ are the ramified 
instanton weights, which will be later identified with the quantities $q_I$ used in
the previous sections (see in particular (\ref{qIdefn}) and (\ref{qMdefn})). The parameters $\epsilon_1$
and $\hat{\epsilon}_2=\epsilon_2/M$ specify the $\Omega$-background \cite{Nekrasov:2002qd,Nekrasov:2003rj} 
which is introduced to localize the integrals over the instanton moduli space; the
rescaling by a factor of $M$ in $\epsilon_2$ is due to the $\mathbb{Z}_M$-orbifold that
is used in the ramified instanton case \cite{Kanno:2011fw}. Finally, the function $g$ is simply
\begin{equation}
g(x)=x~.
\label{g4d}
\end{equation}
This seems an unnecessary redundancy but we have preferred to introduce it because, as we will see later, 
in the five-dimensional theory the integrand of the ramified instanton partition function will have exactly 
the same form as in (\ref{zexplicit5d}), with simply a different function $g$.

The integrations over $\chi_I$ in (\ref{Zso4d5d}) have to be suitably defined and regularized, and we will 
describe this in detail. But first we discuss a few consequences of the integral expression itself and show 
how to extract the twisted chiral superpotential from $Z_{\text{inst}}$.

An immediate feature of (\ref{zexplicit5d}) is that, unlike the case of 
the ${\mathcal N}=2^{\star}$ theory studied in \cite{Ashok:2017odt}, the counting parameters 
$q_I$ have a mass dimension. In order to fix it, let us consider the contribution to the 
partition function coming from the one-instanton sector. This is a sum over 
$M$ terms, each of which has $d_I=1$ for $I=1,\ldots,M$. Explicitly, we have 
\begin{equation}
Z_{1-\text{inst}} = -\sum_{I=1}^M 
 \int \!\frac{d\chi_I}{2\pi \ii}\, \frac{q_I}{\epsilon_1}\prod_{s=1}^{n_I}
\frac{1}{\left(a_{I,s}-\chi_I + \frac 12 (\epsilon_1 + \hat\epsilon_2)\right)}
\prod_{t=1}^{n_{I+1}}\frac{1}{\left(\chi_I 
- a_{I+1,t} + \frac 12 (\epsilon_1 + \hat\epsilon_2)\right)}~.
\end{equation}
Since the partition function is dimensionless and $\chi_I$ carries the dimension of a mass, 
we deduce that mass dimension of $q_I$ is
\begin{equation}
\big[q_I\big] = n_I+n_{I+1}=b_I
\label{massdim}
\end{equation}
where the last step follows from combining (\ref{kI}) and (\ref{bI}).
Another important dimensional constraint follows once we extract the non-perturbative contributions
to the prepotential ${F}$ and to the twisted effective superpotential ${W}$ 
from $Z_{\text{inst}}$. 
This is done by taking the limit in which the $\Omega$-deformation parameters $\epsilon_i$ are set to 
zero according to \cite{Alday:2009fs, Alday:2010vg, Kanno:2011fw}
\begin{equation}
\log Z_{\text{inst}}= -\frac{{F}_{\text{inst}}}{\epsilon_1\hat\epsilon_2}  
+ \frac{{W}_{\text{inst}}}{\epsilon_1} + \ldots
\label{FandW}
\end{equation}
where the ellipses refer to regular terms. The key point is that the prepotential extracted this way depends 
only on the product of all the $q_I$. On the other hand, it is well-known that the instanton contributions 
to the prepotential are organized at weak coupling as a power series expansion in $\Lfour^{2N}$ where
$\Lambda$ is the dynamically generated scale of the four-dimensional theory and $2N$ is the one-loop
coefficient of the gauge coupling $\beta$-function. Thus, we are naturally led to write\,\footnote{The sign
in this formula is the one that, given our conventions, is consistent with the standard field theory results.}
\begin{equation}
\prod_{I=1}^{M} q_I = (-1)^N\Lfour^{2N}~.
\end{equation}
Notice that the mass-dimensions (\ref{massdim}) attributed to each of the $q_I$ are perfectly consistent
with this relation, since the integers $n_I$ form a partition of $N$. 
We therefore find that we can use exactly the same parametrization used in the effective
field theory and given in (\ref{qIdefn}) and (\ref{qMdefn}), which we rewrite here for convenience
\begin{equation}
\begin{aligned}
q_I &=  (-1)^{k_{I-1}}~\Lambda_I^{b_I}\quad\text{for}\quad I = 1, \ldots,  M-1~,\\
q_M&= (-1)^N\,\Lfour^{2N}\Big(\prod_{I=1}^{M-1} q_I\Big)^{-1}~.
\end{aligned}
\label{qIxI}
\end{equation}

\subsubsection*{Residues and contour prescriptions}
The last ingredient we have to specify is how to evaluate the integrals over $\chi_I$ in (\ref{Zso4d5d}). 
The standard prescription \cite{Moore:1998et, Billo':2015ria,Billo':2015jta, Ashok:2017odt} is 
to consider $a_{I,s}$ to be real and then close the integration contours in the upper-half 
$\chi_{I,\sigma}\,$-planes 
with the choice 
\begin{equation}
\mathrm{Im}\, \hat\epsilon_2\gg \mathrm{Im}\, 
\epsilon_1 > 0~.
\label{prescription}
\end{equation}
It is by now well-established that with this prescription the multi-dimensional integrals 
receive contributions from a subset of poles of $z_{\{d_I\}}$, which 
are in one-to-one correspondence with a set of Young diagrams $Y = \{Y_{I,s}\}$, 
with $I=1,\cdots,M$ and $s=1,\cdots n_I$. This fact can be exploited to organize the result in a systematic
way (see for example \cite{Ashok:2017odt} for details).

Let us briefly illustrate this for SU$(2)$, for which there is only one allowed partition, namely $[1,1]$,
and hence one single surface operator to consider \cite{Awata:2010bz}.
 In Tab.~\ref{ZlistSU2} we list the explicit results
for this case, including the location of the poles and the contribution due to all the relevant 
Young tableaux configurations up to two boxes.
\begin{table}[t]
\begin{center}
\begin{tabular}{|c|c|c|c|}
\hline
$\!\!\text{weight}\!\!$ &  poles & $\phantom{\big|}Y$ & $Z_Y$\cr
\hline
$q_1$ & $\phantom{\Big|}\chi_{1,1} = a+ \frac 12\left(\epsilon_1 
+ \hat\epsilon_2\right) $ & $\!\!\left(\Yfund,\bullet\right)\!\!$ & 
$\frac{1}{\epsilon _1 \left(2 a+\epsilon _1+\hat{\epsilon} _2\right)}$ \cr
\hline
$q_2$ & $\phantom{\Big|}\chi_{2,1} = -a+ \frac 12\left(\epsilon_1 
+ \hat\epsilon_2\right)$ & $\!\!\left(\bullet,\Yfund\right)\!\!$ & 
$\frac{1}{\epsilon _1 \left(-2 a+\epsilon _1+\hat{\epsilon} _2\right)}$ \cr
\hline
$q_1 q_2$ & 
\begin{tabular}{@{}c@{}}$\phantom{\Big|}\chi_{1,1} = a+ \frac 12\left(\epsilon_1 
+ \hat\epsilon_2\right) $ \\ $ \phantom{\Big|}\chi_{2,1} = -a + \frac 12\left(\epsilon_1 + \hat\epsilon_2\right)$
\end{tabular}
& $\!\!\left(\Yfund,\Yfund\right)\!\!$ 
&$-\frac{1}{\epsilon _1^2 \left(4 a^2-\hat{\epsilon} _2^2\right)}$ \cr
\hline
$q_1q_2$ & 
\begin{tabular}{@{}c@{}}$\phantom{\Big|}\chi_{1,1} = a + \frac 12\left(\epsilon_1 + \hat\epsilon_2\right)$ \\ 
$\phantom{\Big|}\chi_{2,1} = \chi_{1,1} + \hat\epsilon_2$
\end{tabular}&
$\!\!\left(\Ysymm,\bullet\right)\!\!$ &   
 $-\frac{1}{2 \epsilon _1 \hat{\epsilon} _2 \left(2 a+\hat{\epsilon} _2\right) 
\left(2 a+\epsilon _1+\hat{\epsilon} _2\right)}$ 
\cr
\hline
$q_1 q_2$ & 
\begin{tabular}{@{}c@{}}
 $\phantom{\Big|}\chi_{2,1} =- a + \frac 12\left(\epsilon_1 + \hat\epsilon_2\right)$ 
 \\
 $\phantom{\Big|}\chi_{1,1} = \chi_{2,1} + \hat\epsilon_2$ 
 \end{tabular}&
 $\!\!\left(\bullet,\Ysymm\right)\!\!$ &  
 $-\frac{1}{2 \epsilon _1 \hat{\epsilon} _2 \left(\hat{\epsilon} _2-2 a\right) 
\left(-2 a+\epsilon _1+\hat{\epsilon} _2\right)}$  
\cr
\hline
$q_1^2$ & 
\begin{tabular}{@{}c@{}}
$\phantom{\Big|}\chi_{1,1} =a + \frac 12\left(\epsilon_1 + \hat\epsilon_2\right) $  \\ 
$\phantom{\Big|}\chi_{1,2}=\chi_{1,1}+\epsilon_1$ 
\end{tabular}& 
$\!\!\left(\Yasymm, \bullet\right)\!\!$ & 
$\frac{1}{2 \epsilon _1^2 \left(2 a+\epsilon _1+\hat{\epsilon} _2\right) 
\left(2 a+2 \epsilon _1+\hat{\epsilon} _2\right)}$ \cr
\hline
$q_2^2$ & 
\begin{tabular}{@{}c@{}}
$\phantom{\Big|}\chi_{2,1}=-a + \frac 12\left(\epsilon_1 + \hat\epsilon_2\right) $ \\
$\phantom{\Big|}\chi_{2,2} =\chi_{2,1}+\epsilon_1$ 
\end{tabular}&
$\!\!\left(\bullet, \Yasymm \right)\!\!$ &  
$\frac{1}{2 \epsilon _1^2 \left(-2 a+\epsilon _1+\hat{\epsilon} _2\right) 
\left(-2 a+2 \epsilon _1+\hat{\epsilon} _2\right)}$  \cr
\hline
\end{tabular}
\end{center}
\caption{We list the weight factors, the locations of the poles, the corresponding Young 
diagrams, and the contribution to the partition function in all cases up to two boxes for the SU(2) theory. 
Here we have set $a_1=-a_2=a$.} 
\label{ZlistSU2}
\end{table}

Combining these results, we find that the instanton partition function takes the following form 
\begin{align}
Z_{\text{inst}}[1,1] &= 1+\frac{q_1}{\epsilon _1 \left(2 a+\epsilon _1+\hat{\epsilon} _2\right)} 
+ \frac{q_2}{\epsilon _1 \left(-2 a+\epsilon _1+\hat{\epsilon} _2\right)}\notag\\
&~+\frac{q_1^2}{2 \epsilon _1^2 \left(2 a+\epsilon _1+\hat{\epsilon} _2\right) 
\left(2 a+2 \epsilon _1+\hat{\epsilon} _2\right)}+\frac{q_2^2}{2 \epsilon _1^2 
\left(-2 a+\epsilon _1+\hat{\epsilon} _2\right) \left(-2 a+2 \epsilon _1+\hat{\epsilon} _2\right)} 
\notag\\
&~+q_1q_2\,\frac{\epsilon _1+\hat\epsilon _2}{\epsilon _1^2 \hat\epsilon _2 
\left(-2 a+\epsilon _1+\hat\epsilon _2\right) \left(2 a+\epsilon _1+\hat\epsilon _2\right)}+ \ldots\label{ZinstSU2}
\end{align}
The prepotential and the twisted effective superpotential are extracted according to \eqref{FandW} and
using the map (\ref{qIxI}). Let us focus on the twisted superpotential ${W}_{\text{inst}}$, or better
on its $q_1$-derivative. We find
\begin{equation}
\begin{aligned}
q_1 \frac{d {W}_{\text{inst}}}{d q_1}
= \frac{1}{2a}\Big(\Lambda_1^2 &\,+ \frac{\Lfour^4}{\Lambda_1^2} \Big) 
-\frac{1}{8a^3}\Big(\Lambda_1^4 + \frac{\Lfour^8}{\Lambda_1^4}\Big) 
+\ldots~.
\end{aligned}
\label{Wsu2loc}
\end{equation}
This precisely matches, up to two instantons, the non-perturbative part of the result (\ref{Wsu2TCR}) 
obtained by solving 
the twisted chiral ring equations for the quiver theory representing the surface defect in SU(2).
We have also checked the agreement at higher instanton orders (up to six boxes), which we have not reported
here for brevity.

The specific prescription (\ref{prescription}) we have chosen to compute the instanton partition function
is particularly nice due to the correspondence of the residues with Young tableaux. However, 
there are many other possible choices of contours that one can make. One way to classify these 
distinct contours is using the Jeffrey-Kirwan (JK) prescription \cite{JK1995}. In this terminology, 
the set of poles chosen to compute the
residues is described by a JK parameter $\eta$, which is a particular 
linear combination of the $\chi_{I,s}$; the prescription chooses a set of factors ${\mathcal D}$ 
from the denominator of $z_{\{d_I\}}$ such that, if we only consider the $\chi_{I,s}$-dependent terms of 
these chosen factors, then, $\eta$ can be written as a positive linear combination of these. 
For instance, our prescription in (\ref{prescription}) corresponds to choosing\,\footnote{We 
understand the extra index $s$ running from $1$ to $n_I$.}
\begin{equation}
\eta=-\sum_{I=1}^M\chi_{I}
\label{JK1}
\end{equation}
For a detailed discussion of this method in the context of ramified instantons we refer
to \cite{Gorsky:2017hro} where it is also shown that different JK prescriptions can be mapped to 
different quiver realizations of the surface operator. 

Let us consider for example the prescription corresponding to a JK parameter of the form
\begin{equation}
\eta=-\sum_{I=1}^{M-1}\chi_{I}+ \zeta\,\chi_M
\label{JK2}
\end{equation}
where $\zeta$ is a large positive number. In our notation this corresponds to closing the integration
contours in the upper half-plane as before for the first $(M-1)$ variables, and in the lower half plane
for $\chi_M$. Applying this new prescription to the SU(2) theory, we find a
different set of poles that contribute. They are explicitly listed in Tab.~\ref{ZlistSU2new}.
\begin{table}[t]
\begin{center}
\begin{tabular}{|c|c|c|}
\hline
 $\!\!\text{weight}\!\!$ &  poles & $Z_Y$\cr
\hline
 $q_1$ & 
$\phantom{\Big|}\chi_{1,1} = a+ \frac 12\left(\epsilon_1 
+ \hat\epsilon_2\right) $ & $\frac{1}{\epsilon _1 \left(2 a+\epsilon _1+\hat{\epsilon} _2\right)}$ \cr
\hline
  $q_2$ & 
$\phantom{\Big|}\chi_{2,1} = a- \frac 12\left(\epsilon_1 
+ \hat\epsilon_2\right)$ & $\frac{1}{\epsilon _1 \left(-2 a+\epsilon _1+\hat{\epsilon} _2\right)}$ \cr
\hline
$q_1q_2$ & 
\begin{tabular}{@{}c@{}}$\phantom{\Big|}\chi_{1,1} = \chi_{2,1}+\hat\epsilon_2$ \\ 
$\phantom{\Big|}\chi_{2,1} = -a-\frac{1}{2}(\epsilon_1 +3 \hat\epsilon_2)$
\end{tabular}
 &  $-\frac{1}{2 \epsilon _1 \hat\epsilon _2 \left(2 a+\hat\epsilon _2\right) 
\left(2 a+\epsilon _1+\hat\epsilon _2\right)}$ 
\cr
\hline
 $q_1 q_2$ & 
\begin{tabular}{@{}c@{}}
 $\phantom{\Big|}\chi_{1,1} = \chi_{2,1}+\hat\epsilon_2$ 
 \\
 $\phantom{\Big|}\chi_{2,1} = a-\frac{1}{2}(\epsilon_1+\hat\epsilon_2)$ 
 \end{tabular}
 & $\frac{\epsilon _1+2 \hat\epsilon _2}{2 \epsilon _1^2 \hat\epsilon _2 
\left(2 a+\hat\epsilon _2\right) \left(-2 a+\epsilon _1+\hat\epsilon _2\right)}$  
\cr
\hline
 $q_1^2$ & 
\begin{tabular}{@{}c@{}}
$\phantom{\Big|}\chi_{1,1} =a_{1} + \frac 12\left(\epsilon_1 + \hat\epsilon_2\right) $  \\ 
$\phantom{\Big|}\chi_{1,2}=\chi_{1,1}+\epsilon_1$ 
\end{tabular}
& $\frac{1}{2 \epsilon _1^2 \left(2 a+\epsilon _1+\hat{\epsilon} _2\right) 
\left(2 a+2 \epsilon _1+\hat{\epsilon} _2\right)}$ \cr
\hline
 $q_2^2$ & 
\begin{tabular}{@{}c@{}}
$\phantom{\Big|}\chi_{2,1}=a - \frac 12\left(\epsilon_1 + \hat\epsilon_2\right) $ \\
$\phantom{\Big|}\chi_{2,2} =\chi_{2,1}-\epsilon_1$ 
\end{tabular}
& $\frac{1}{2 \epsilon _1^2 \left(-2 a+\epsilon _1+\hat{\epsilon} _2\right) 
\left(-2 a+2 \epsilon _1+\hat{\epsilon} _2\right)}$  \cr
\hline
\end{tabular}
\end{center}
\caption{We list the weight factors, the pole structure and the contribution 
to the partition function in all cases up to two boxes for the SU(2) theory using 
the contour prescription corresponding to the
JK parameter (\ref{JK2}).} 
\label{ZlistSU2new}
\end{table}

Comparing with Tab.~\ref{ZlistSU2}, we see that, although the location of residues has changed, for 
most cases the residues are unchanged. The only set of residues that give an apparently different 
answer is the one with $d_1=d_2=1$ with weight $q_1q_2$. As opposed to the earlier case, 
where there were three contributions, now there are only two terms proportional to $q_1q_2$. 
However, it is easy to see that if we sum these contributions, we find an exact match between the two prescriptions.
This fact should not come as surprise since it is a simple consequence of the residue theorem applied
to the $\chi_2$ integral. Therefore, all results that follow from the instanton partition function 
(and in particular the twisted superpotential) are the same in the two cases.
Of course what we have just seen in the simple SU(2) case at the two instanton level,
occurs also at higher instanton numbers and with higher rank gauge groups.
The price one pays in changing the contour prescription or equivalently in changing the JK parameter
from (\ref{JK1}) to (\ref{JK2}) is the loss of a simple one-to-one correspondence with 
the Young tableaux, but the gain is that, as shown in \cite{Gorsky:2017hro}, the second prescription 
produces at each instanton order an instanton partition that is already organized in a factorized form 
in which the various factors account for the 2d, the 4d and the mixed 2d/4d contributions. 
This is a feature that will play a fundamental role in the 3d/5d extension. 

Let us now list our findings obtained by using the second residue prescription for the SU(3) theory, 
limiting ourselves to the one-instanton terms for brevity. 
In the case of the surface operator corresponding to the partition [1,2] we get 
\begin{equation}
\begin{aligned}
Z_{\text{inst}}[1,2] =&\,1+
\frac{q_1}{\epsilon _1 \left(a_{12}+\epsilon _1+\hat{\epsilon} _2\right)
\left(a_{13}+\epsilon _1+\hat{\epsilon} _2\right)} 
\!+\! \frac{q_2}{\epsilon _1 \left(a_{21}+\epsilon _1+\hat{\epsilon} _2\right)
\left(a_{31}+\epsilon _1+\hat{\epsilon} _2\right)}+\ldots
\end{aligned}
\label{ZSU312}
\end{equation}
while for the surface operator described by the partition [1,1,1] we obtain
\begin{equation}
\begin{aligned}
Z_{\text{inst}}[1,1,1] =&\,1+
\frac{q_1}{\epsilon _1 \left(a_{12}+\epsilon _1+\hat{\epsilon} _2\right)} +
\frac{q_2}{\epsilon _1 \left(a_{23}+\epsilon _1+\hat{\epsilon} _2\right)} 
+\frac{q_3}{\epsilon _1 \left(a_{31}+\epsilon _1+\hat{\epsilon} _2\right)}+\ldots
~.
\end{aligned}
\label{ZSU3111}
\end{equation}
Applying (\ref{FandW}) to extract $W_{\text{inst}}$, we 
find that the $q_I$-logarithmic derivatives of the twisted superpotential for the two partitions perfectly match 
the non-perturbative pieces of the solutions
(\ref{sigma1}) and (\ref{sigma12}) of the twisted chiral ring equations.
We have checked that this agreement persists at the two-instanton level. 
We have also thoroughly explored all surface operators in the SU(4) theory and many cases
in higher rank theories up to two instantons, always finding a perfect match between the
$q_I$-logarithmic derivatives of ${W}$ and the solutions of the corresponding
twisted chiral ring equations.

\subsection{Localization in 5d}
We now turn to discuss the results for a gauge theory on $\mathbb{R}^4\times S_\beta^1$
in the presence of a surface operator also wrapping the compactification circle.
This case has been discussed by a number of recent works (see for instance 
\cite{Gaiotto:2014ina,Bullimore:2014awa}).

Here we observe that the ramified instanton partition function is given by the same expressions
(\ref{Zso4d5d}) and (\ref{zexplicit5d}) in which the function $g(x)$ is 
\cite{Nekrasov:2002qd,Nekrasov:2003rj,Hollowood:2003cv}
\begin{equation}
g(x)=2\sinh\frac{\beta x}{2}
\label{g5d}
\end{equation}
Another difference with respect to the 2d/4d case is that the counting parameters
$q_I$ are now dimensionless and
are given by
\begin{equation}
\begin{aligned}
q_I &=  (-1)^{k_{I-1}}~\big(\beta\Lambda_I\big)^{b_I}\quad\text{for}
\quad I = 1, \ldots,  M-1~,\\
q_M&= (-1)^N\,\big(\beta\Lfour\big)^{2N}\Big(\prod_{I=1}^{M-1} q_I\Big)^{-1}~.
\end{aligned}
\label{qIxI5d}
\end{equation}
The final result is obtained by summing the residues of $z_{\{d_I\}}$ over the same set of poles selected
by the JK prescription (\ref{JK2}). 

Let us illustrate these ideas by calculating the twisted effective superpotential that governs the 
infrared behavior of the $[1,1]$ operator in SU$(2)$. Up to two instantons, the partition function 
using these rules is
\begin{align}
Z_{\text{inst.}}[1,1]=& \,1
+\frac{q_1}{4\sinh \big(\frac{\beta}{2}  \epsilon_1\big) 
\sinh \big(\frac{\beta}{2}(2a+\epsilon_1+\hat{\epsilon}_2)\big)} 
+\frac{q_2}{4\sinh \big(\frac{\beta}{2} \epsilon_1\big) 
\sinh \big(\frac{\beta}{2}(-2a+\epsilon_1+\hat{\epsilon}_2)\big)} \notag\\
&+ \frac{q_1^2}{16\sinh\big(\frac{\beta}{2}\epsilon_1\big) \sinh\big(\beta  \epsilon_1\big) 
\sinh\big(\frac{\beta}{2}(2a+\epsilon_1+\hat{\epsilon}_2)\big)\sinh\big(\frac{\beta}{2}
(2a +2\epsilon_1+\hat{\epsilon}_2)\big)} \notag\\
&+ \frac{q_2^2}{16\sinh\big(\frac{\beta}{2}\epsilon_1\big) \sinh\big(\beta  \epsilon_1\big) 
\sinh\big(\frac{\beta}{2}(-2a+\epsilon_1+\hat{\epsilon}_2)\big)\sinh\big(\frac{\beta}{2}
(-2a +2\epsilon_1+\hat{\epsilon}_2)\big)} \notag\\
&+ \frac{q_1q_2\,\sinh\big(\frac{\beta}{2}(\epsilon_1+2\hat{\epsilon}_2)\big)}{16
\sinh^2\!\big(\frac{\beta}{2}\epsilon_1\big)\sinh\big(\beta\hat{\epsilon_2}\big)
\sinh\big(\frac{\beta}{2}(-2a+\epsilon_1+\hat{\epsilon}_2)\big)
\sinh\big(\frac{\beta}{2}(2a+\hat{\epsilon}_2)\big)}
\notag\\
&+\frac{q_1q_2\phantom{\big|}}{16 \sinh\big(\frac{\beta}{2}\epsilon_1\big)\sinh\big(\beta\hat{\epsilon}_2\big)
\sinh\big(\frac{\beta}{2}(2a+\epsilon_1+\hat{\epsilon}_2)\big)
\sinh\big(\frac{\beta}{2}(2a+\hat{\epsilon}_2)\big)}\notag\\
&+\ldots\phantom{\Big|}
\end{align}
where $a_1=-a_2=a$. {From} this instanton partition function we can extract the twisted chiral superpotential
in the usual manner according to (\ref{FandW}).  The result is 
\begin{align}
\label{WlocSU2b}
q_1 \frac{d W_{\text{inst}}}{d q_1}  
&= \frac{\beta}{2\sinh(\beta a)}\Big(\Lambda_1^2+\frac{\Lfour^4}{\Lambda_1^2}
\Big)-\frac{\beta^3\cosh(\beta a)}{8\sinh^3(\beta a)}
\Big(\Lambda_1^4+\frac{\Lfour^8}{\Lambda_1^4}\Big)+\ldots
\end{align}
It is very easy to check that in the limit $\beta\to 0$ this expression reduces to the 2d/4d result
in (\ref{Wsu2loc}). Most importantly it agrees with the non-perturbative part of the solution
(\ref{sigmasol}) of the chiral ring equation of the 3d/5d SU(2) theory, thus confirming the validity
of (\ref{derWsol}).

Similar calculations can be performed for the higher rank cases without much difficulty, and indeed we 
have done these calculations for all surface operators of SU(4) and for many cases up to SU(6). 
Here, for brevity, we simply report the results at the one-instanton level for the surface operators 
in the SU(3) theory. In the case of the defect of type [1,2] the instanton partition function is
\begin{align}
Z_{\mathrm{inst}}[1,2]=&\,1+\frac{q_1}{8\sinh \big(\frac{\beta}{2}  \epsilon_1\big) 
\sinh \big(\frac{\beta}{2}(a_{12}+\epsilon_1+\hat{\epsilon}_2)\big)
\sinh \big(\frac{\beta}{2}(a_{13}+\epsilon_1+\hat{\epsilon}_2)\big)}\label{Zsu3125d} \\
&+
\frac{q_2}{8\sinh \big(\frac{\beta}{2}  \epsilon_1\big) 
\sinh \big(\frac{\beta}{2}(-a_{12}+\epsilon_1+\hat{\epsilon}_2)\big)
\sinh \big(\frac{\beta}{2}(-a_{13}+\epsilon_1+\hat{\epsilon}_2)\big)} +\ldots~,\notag
\end{align}
while for the defect of type [1,1,1] we find
\begin{align}
Z_{\mathrm{inst}}[1,1,1]=&\,1+\!\frac{q_1}{4\sinh \big(\frac{\beta}{2}  \epsilon_1\big) 
\sinh \big(\frac{\beta}{2}(a_{12}+\epsilon_1+\hat{\epsilon}_2)\big)}
+\!\frac{q_2}{4\sinh \big(\frac{\beta}{2}  \epsilon_1\big) 
\sinh \big(\frac{\beta}{2}(a_{23}+\epsilon_1+\hat{\epsilon}_2)\big)}\notag
 \\
 &+\frac{q_3}{4\sinh \big(\frac{\beta}{2}  \epsilon_1\big) 
\sinh \big(\frac{\beta}{2}(a_{31}+\epsilon_1+\hat{\epsilon}_2)\big)}+\ldots
\label{Zsu31115d}
\end{align}
where $a_{ij}=a_1-a_j$. These expressions are clear generalizations of the 2d/4d instanton partition functions
(\ref{ZSU312}) and (\ref{ZSU3111}). Moreover one can check that the twisted superpotentials that can
be derived from them perfectly match the ones obtained by solving the chiral ring equations as we discussed in
Section~\ref{quiver3d5d}.

\section{Superpotentials for dual quivers}
\label{secn:dual}
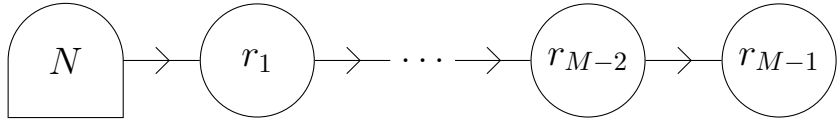
\begin{figure}[ht]
\centering
\begin{tikzpicture}[decoration={
markings,% switch on markings
mark=at position 0.6 with {\draw (-5pt,-5pt) -- (0pt,0pt);
                \draw (-5pt,5pt) -- (0pt,0pt);}}]
  \matrix[row sep=10mm,column sep=5mm] {
       \node(gfN)[gaugedflavor]{\Large $N$}; & & 
       \node(g2)[gauge] {\Large $r_{1}$};  && 
       \node(dots){\Large $\ldots$}; & & 
       \node(glast)[gauge] {\Large $r_{M-2}$};& &
      \node(g1)[gauge] {\Large $r_{M-1}$}; \\
  };
\graph{(gfN) --[postaction={decorate}](g2) --[postaction={decorate}](dots)--[postaction={decorate}](glast) 
--[postaction={decorate}](g1);};
\end{tikzpicture}
\label{dualquiverpicgeneric1}
\vspace{-0.2cm}
\caption{The quiver which is dual to the one in Fig-~\ref{quiverpicgeneric1}.}
\end{figure}

The 2d/4d quiver theories considered in Section~\ref{quiver2d4d} admit dual 
descriptions \cite{Benini:2014mia,Closset:2015rna,Gorsky:2017hro}. In particular, 
with repeated applications of Seiberg-like dualities, one can prove that the linear quiver of 
Fig.~\ref{quiverpicgeneric1} is dual to the one represented in Fig.~\ref{dualquiverpicgeneric1}.
Here the ranks of the U($r_I$) gauge groups are given by 
\begin{equation}
\label{dualr}
r_I = N-k_I = \sum_{K=I+1}^{M} n_{K}~,
\end{equation}
where in the second step we have used (\ref{kI}) to express $k_I$ in terms of the entries of partition 
$[n_1,\ldots,n_M]$ labeling the surface defect. Notice the reversal of the arrows  with respect 
to the quiver in Fig.~\ref{quiverpicgeneric1}, and thus the different assignment of massive chiral 
fields to fundamental or anti-fundamental representations. 

The new quiver provides an alternative realization of the same surface operator \cite{Gorsky:2017hro}.
Its corresponding twisted superpotential, which we denote by $\widetilde{W}$, is given by the obvious 
modification of (\ref{tildeW1}), and reads
\footnote{For later convenience, we denote the twisted chiral scalars and the FI couplings 
of the dual gauge groups by tilde variables.}
\begin{equation}
\begin{aligned}
\widetilde{W} =& ~2\pi\ii \sum_{I=1}^{M-1}\sum_{s=1}^{r_I}\widetilde\tau_I (\mu)\,
\widetilde\sigma^{(I)}_s -\!
\sum_{I=2}^{M-1} \sum_{s=1}^{r_{I}} \sum_{t=1}^{r_{I-1}}
\varpi\big(\widetilde\sigma^{(I-1)}_t - \widetilde\sigma^{(I)}_s\big) \\
&\qquad-\!\sum_{s=1}^{r_1}\Big\langle 
\Tr \varpi\big(\Phi - \widetilde\sigma^{(1)}_s ) \Big\rangle~.
\end{aligned}
\label{tildeW1d}
\end{equation}
As in (\ref{tildeW1}), the linear terms in $\widetilde{\tau}_I(\mu)$ are the classical contributions, while the
other terms are the one-loop part. 
The dual FI couplings $\widetilde\tau_I(\mu)$ renormalize like the orginal couplings $\tau_I(\mu)$ but with 
$k_I$ replaced by $r_I$. In view of (\ref{dualr}), this implies that the one-loop $\beta$-function
coefficient in the dual theory is opposite to that of the original theory, namely
\begin{equation}
\widetilde{b}_I=r_{I+1}-r_{I-1}=-k_{I+1}+k_{I-1}=-b_I~.
\label{dualbI}
\end{equation}
In turn, this implies that the dynamically generated scale in the $I$-th node of the dual theory is
given by
\begin{equation}
\widetilde{\Lambda}_I^{b_I}=\rme^{-2\pi\ii\,\widetilde{\tau}_I}\,\mu^{b_I}~,
\label{dualLambdaI}
\end{equation}
to be compared with (\ref{defLambdaI}). As usual we can trade 
the couplings $\widetilde{\tau}_I(\mu)$ for these
scales $\widetilde{\Lambda}_I$, and thus rewrite the twisted superpotential (\ref{tildeW1d}) in a form that
is the straightforward modification of (\ref{WHat}).

If we make the following classical ansatz
\begin{equation}
\label{ansd}
\widetilde\sigma^{(I)} = \mathrm{diag}(a_{n_1 + \ldots + n_{I}+1}, 
a_{n_1 + \ldots + n_{I} + 2}, \ldots, a_N)~,
\end{equation}
which is dual to the one for $\sigma^{(I)}$ given in (\ref{sigmavac}), then
it is easy to check that
\begin{equation}
\label{relsds}
\Tr \widetilde\sigma^{(I)} = -\Tr \sigma^{(I)}~.
\end{equation}
This clearly implies
\begin{equation}
\frac{1}{2\pi\ii}\frac{\partial \widetilde W_{\mathrm{class}}}{\partial \widetilde \tau_I} = 
- \frac{1}{2\pi\ii} \frac{\partial W_{\mathrm{class}}}{\partial \tau_I}~.
\label{xWtilde}
\end{equation}
Thus, if the FI parameters in the two dual models are related to each other by
\begin{equation}
\label{WcWtc}
\widetilde\tau_I = -\tau_I~,
\end{equation}
one has $\widetilde W_{\mathrm{class}}=W_{\mathrm{class}}$. Notice that using (\ref{WcWtc})
in (\ref{dualLambdaI}) and comparing with (\ref{defLambdaI}), we have
\begin{equation}
\widetilde{\Lambda}_I=\Lambda_I~.
\label{dLaisLa}
\end{equation}

The relation (\ref{relsds}) remains true also at the quantum level. This statement can be verified 
by expanding $\widetilde{\sigma}^{(I)}$ as a power series in the various $\widetilde{\Lambda}_I$’s around 
the classical vacuum (\ref{ansd}), and iteratively solving the corresponding chiral ring 
equations in a semi-classical approximation. Doing this and using (\ref{WcWtc}) and (\ref{dLaisLa}), 
we have checked the validity of (\ref{relsds}) in several examples.
Furthermore, we have obtained the same relations also using the localization methods 
described in Section~\ref{localize4d5d}. Therefore,
we can conclude that the two quiver theories in Fig.~\ref{quiverpicgeneric1} and~\ref{dualquiverpicgeneric1},
indeed provide equivalent descriptions of the 2d/4d defect SU$(N)[n_1,\ldots,n_M]$.

This conclusion changes drastically once we consider the 3d/5d quiver theories compactified 
on a circle. In this case, the dual superpotential
corresponding to the quiver in Fig.~\ref{dualquiverpicgeneric1}, is obtained by upgrading 
(\ref{tildeW1d}) to a form analogous to (\ref{tildeW15d}), namely
\begin{equation}
\begin{aligned}
\widetilde{W}=&\sum_{I=1}^{M-1}\sum_{s=1}^{r_I}
\widetilde{b}_I\,\log (\beta\widetilde{\Lambda}_I)\widetilde\sigma^{(I)}_s
- \sum_{I=2}^{M-1} \sum_{s=1}^{r_{I}} \sum_{t=1}^{r_{I-1}}
\ell\big(\widetilde\sigma^{(I-1)}_t - \widetilde\sigma^{(I)}_s\big)
-\sum_{s=1}^{r_1}\Big\langle 
\Tr \ell\big(\Phi - \widetilde\sigma^{(1)}_s \big) \Big\rangle~.
\end{aligned}
\label{dtildeW15d}
\end{equation}
Here we have used the loop-function $\ell(x)$ defined in (\ref{defell}), and taken into 
account the renormalization of the FI couplings to introduce the scales $\widetilde{\Lambda}_I$. Using for 
the original quiver the ansatz (\ref{sigmavace}), and for the dual theory the ansatz (\ref{ansd}), 
which can be rewritten as 
\begin{equation}
\label{ansde}
\widetilde{S}^{(I)} = \mathrm{diag}(A_{n_1 + \ldots + n_{I + 1}}, A_{n_1 + \ldots + n_{I + 2}}, \ldots, A_N)
\end{equation}   
in terms of the exponential variables $\widetilde{S}^{(I)}=\rme^{\beta \widetilde{\sigma}^{(I)}}$ 
and $A_i=\rme^{\beta a_i}$, one can easily check that
the relation (\ref{relsds}) still holds true.

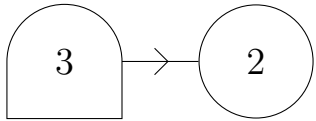
\begin{figure}[ht]
\begin{center}
\begin{tikzpicture}[decoration={
markings,% switch on markings
mark=at position 0.6 with {\draw (0pt,0pt) -- (-5pt,-5pt);
                \draw (0pt,0pt) -- (-5pt,5pt);}}]
  \matrix[row sep=10mm,column sep=5mm] {
       \node(gf2)[gaugedflavor]{\Large $3$};    & &\node(g1)[gauge] {\Large $2$}; \\
  };
\graph{(gf2) --[postaction={decorate}](g1);};
\end{tikzpicture}
\end{center}
\vspace{-0.5cm}
\caption{The dual quiver for the SU$(3)[1,2]$ defect.}
\label{fig:su312dq}
\end{figure}

However, in general, this is no longer valid for the full solutions of the chiral ring equations. 
This happens whenever the ranks $k_I$ of the original quiver theory
and the ranks $r_I$ of the dual model are different from each other for some $I$, which is the 
generic situation.
Let us show this in a specific example, namely the defect of type [1,2] in the SU(3) theory.
The original quiver theory was discussed in detail in Section~\ref{quiver3d5d} where we have shown that
the solution of the chiral ring equation is (see (\ref{dW135d}))
\begin{equation}
\label{dW135dbis}
\sigma_\star = a_1 + \beta^2 \frac{A_1^{{1}/{2}}}{A_{12}A_{13}}
\Big(\Lambda_1^3+\frac{\Lambda^6}{\Lambda_1^3} \Big)+\ldots~.
\end{equation}
The dual quiver for this defect is depicted in Fig.~\ref{fig:su312dq}. 

\noindent
{From} (\ref{dtildeW15d}), it follows
that the corresponding twisted superpotential is
\begin{equation}
\label{wtilde312}
\widetilde{W}=\sum_{s=1}^2\bigg[
\!- \frac{3}{2\pi\ii}\log (\beta\Lambda_1)
\widetilde{\sigma}_s 
-\Big\langle\Tr \ell(\Phi-\widetilde{\sigma}_s)\Big\rangle\bigg]~.
\end{equation}
Using the function $\widehat{P}_3$ defined in (\ref{P35d}), we see that
the twisted chiral ring equations are 
\begin{equation}
\label{tCr3312}
\widehat{P}_3(\widetilde{S}_s) = \beta^3 \Big(\widetilde{\Lambda}_1^3
+ \frac{\Lfour^6}{\widetilde{\Lambda}_1^3}\Big)
\end{equation} 
for $s=1,2$. Solving iteratively these equations around the classical vacuum (\ref{ansd}), we find
\begin{equation}
\label{soldsigma}
\Tr \widetilde{\sigma}_{\star}=
\,a_2+a_3+\beta^2
\Big(\widetilde{\Lambda}_1^3
+ \frac{\Lfour^6}{\widetilde{\Lambda}_1^3}\Big)
\Big(\frac{A_3^{1/2}}{A_{13}A_{23}}-
\frac{A_2^{1/2}}{A_{12}A_{23}}\Big)
+\ldots~.
\end{equation}
By comparing (\ref{dW135dbis}) and (\ref{soldsigma}), we see that at the classical level
$\Tr \widetilde{\sigma}_{\star}$
is equal to negative of the solution $\sigma_\star$ in the original quiver; this simply follows from the 
SU$(3)$ tracelessness condition. However, the first semi-classical correction of order $\beta^2$ 
spoils this relation, even if we use the relation (\ref{dLaisLa}) between the dynamically generated scales. Therefore, as anticipated, the two descriptions are not any more dual to each other.

It is interesting to observe that the twisted superpotential corresponding to the dual solution (\ref{soldsigma})
can also be obtained using localization. Indeed, if one evaluates the instanton partition 
function $Z_{\mathrm{inst}}[1,2]$ for the compactified theory using the JK prescription with
\begin{equation}
\eta= + \chi_1-\xi\,\chi_2~,
\label{etadual}
\end{equation}
where $\xi$ is positive and large, and then extracts from it the corresponding twisted 
superpotential using (\ref{FandW}), one obtains precisely the above result\,\footnote{We have checked this 
up to the two-instanton level, namely up to order $\beta^5$.}. 
Notice that the JK parameter (\ref{etadual}) is opposite in sign with respect to the one in (\ref{JK2}) 
that we have adopted in the original quiver realization. Actually, what we have seen 
in this particular example can be generalized to other cases and for any $M$, we find that the 
JK parameter which has to be used in the localization computations for the dual quiver theory 
to match the solution of the chiral ring equations is
\begin{equation}
\eta=\sum_{I=1}^{M-1}\chi_I-\xi\,\chi_M~.
\label{etadual1}
\end{equation}
This fact points towards the nice scenario in which
the twisted superpotentials $W$ and $\widetilde{W}$ for a pair of quiver theories
related by a chain of Seiberg-like dualities can be obtained in localization using two different 
JK prescriptions associated to opposite $\eta$ parameters.
While in the 2d/4d systems all different JK prescriptions are equivalent to each other and 
lead to the same superpotentials, in general this is no longer true in the 3d/5d theories because of 
the particular structure of the instanton partition functions.

\subsection{Adding Chern-Simons terms}
We now investigate the possibility of restoring the duality between the two 3d/5d descriptions of the 
SU(3)[1,2] defect by considering the addition of Chern-Simons (CS) couplings. 
These can be written as a term in the twisted chiral superpotential that is quadratic in the 
twisted scalars and proportional to the compactification circle $\beta$ \cite{Nekrasov:2009uh, Chen:2012we}. 
For the $I$th node, the CS term is of the form:
\begin{equation}
W_{\text{CS}} = \frac{k}{2}\,\beta \,\Tr (\sigma^{(I)})^2 ~.
\end{equation}

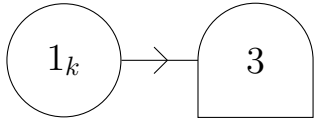
\begin{figure}[t]
\begin{center}
\begin{tikzpicture}[decoration={
markings,% switch on markings
mark=at position 0.6 with {\draw (-5pt,-5pt) -- (0pt,0pt);
                \draw (-5pt,5pt) -- (0pt,0pt);}}]
  \matrix[row sep=10mm,column sep=5mm] {
      \node(g1)[gauge] {\Large $1_k$};      & &\node(gf2)[gaugedflavor]{\Large
      $3$};\\
  };
\graph{(g1) --[postaction={decorate}](gf2);};
\end{tikzpicture}
\end{center}
\vspace{-0.5cm}
\caption{The quiver for the SU(3)[1,2] theory with a CS term with coupling $k$ on the U(1) node.}
\label{quiversu312k}
\end{figure}

Let us start from the original theory and let us turn on a CS term on the U(1) node with coupling $k$. 
The resulting quiver is represented in Fig.~\ref{quiversu312k} and the corresponding
twisted superpotential is
\begin{equation}
W= 3\log(\beta\Lambda_1)\,\sigma
+\frac{k}{2}\,\beta\,\sigma^2-\Big \langle\Tr \ell(\sigma-\Phi)\Big\rangle~.
\label{WCS1}
\end{equation}
Repeating the same steps described in Section~\ref{quiver3d5d}, we easily obtain the modified
twisted chiral ring equation
\begin{equation}
\widehat{P}_3(S) = \beta^3 \Big( S^k \Lambda_1^3 + \frac{ \Lfour^6}{S^k \Lambda_1^3}\Big)
\label{CRCS1}
\end{equation} 
where, as before, $\widehat{P}_3$ is given in (\ref{P35d}) and $S=\rme^{\beta\sigma}$.
Solving this equation with the usual ansatz leads to
\begin{equation}
\sigma_\star =a_1 + \beta^2 \frac{A_1^{{1}/{2}}}{A_{12}A_{13}}
\Big(A_1^k\,\Lambda_1^3+\frac{\Lambda^6}{A_1^k\,\Lambda_1^3} \Big)+\ldots~.
\end{equation}
Of course, for $k=0$ one recovers the solution (\ref{dW135dbis}) in the absence of the CS term.

Let us now consider the dual quiver with a CS interaction with coupling $\widetilde{k}$ turned on in the U(2) node.
\begin{figure}[ht]
\begin{center}
\begin{tikzpicture}[decoration={
markings,% switch on markings
mark=at position 0.6 with {\draw (0pt,0pt) -- (-5pt,-5pt);
                \draw (0pt,0pt) -- (-5pt,5pt);}}]
  \matrix[row sep=10mm,column sep=5mm] {
       \node(gf2)[gaugedflavor]{\Large $3$};    & &\node(g1)[gauge] {\Large $2_{\widetilde{k}}$}; \\
  };
\graph{(gf2) --[postaction={decorate}](g1);};
\end{tikzpicture}
\end{center}
\vspace{-0.5cm}
\caption{The quiver representing the dual realization of the SU(3)[1,2] with a CS term on the U(2) node}
\label{dualSU312k}
\end{figure}
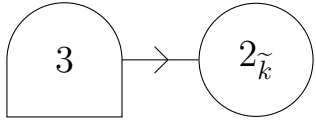

This is represented in Fig.~\ref{dualSU312k} and the corresponding twisted superpotential is
\begin{equation}
\widetilde{W}=\sum_{s=1}^2\bigg[\!- 3\log (\beta\Lambda_1)\,
\widetilde{\sigma}_s 
+\frac{\widetilde{k}}{2}\,\beta\,\widetilde{\sigma}_s^2 
-\Big\langle\Tr \ell(\Phi-\widetilde{\sigma}_s)\Big\rangle\bigg]~.
\label{dualWk}
\end{equation}
{From} this we can easily derive the twisted chiral ring equations, namely
\begin{equation}
\widehat{P}_3(\widetilde{S}_s) = \beta^3 \Big( \frac{\widetilde{\Lambda}_1^3}{\widetilde{S}_s^k} 
+ \frac{ \widetilde{S}_s^k\,\Lfour^6}{\widetilde{\Lambda}_1^3}\Big)
\end{equation}
for $s=1,2$, which are a simple generalization of (\ref{tCr3312}). Solving these equations 
with the ansatz (\ref{ansd}) we find
\begin{equation}
\begin{aligned}
\widetilde{\sigma}_{\star,1}+\widetilde{\sigma}_{\star,2}=&\,a_2+a_3
+\beta^2\bigg[\Big(\frac{A_3^{1/2-\widetilde{k}}}{A_{13}A_{23}}-
\frac{A_2^{1/2-\widetilde{k}}}{A_{12}A_{23}}\Big)\widetilde{\Lambda}_1^3
\!+\!\Big(\frac{A_3^{1/2+\widetilde{k}}}{A_{13}A_{23}}-
\frac{A_2^{1/2+\widetilde{k}}}{A_{12}A_{23}}\Big)\frac{\Lambda^6}{\widetilde{\Lambda}_1^3}
\bigg]+\ldots~.
\end{aligned}
\end{equation}
Of course for $\widetilde{k}=0$ we recover the solution (\ref{soldsigma}) in the absence of the CS coupling.

Our main observation is that if we take
\begin{equation}
k=-\widetilde{k}=\frac{1}{2}\quad
\end{equation}
then, using the SU(3) tracelessness condition and the relations (\ref{WcWtc}) and (\ref{dLaisLa}), we have
\begin{equation}
\widetilde{\sigma}_{\star,1}+\widetilde{\sigma}_{\star,2}=-\sigma_\star
\end{equation}
This implies that
\begin{equation}
\widetilde{\Lambda}_1\frac{d \widetilde W}{d\widetilde{\Lambda}_1}
= -{\Lambda}_1\frac{d W}{d{\Lambda}_1}~,
\end{equation}
so that the duality between $W$ and $\widetilde{W}$ is restored at the quantum level under the map 
 (\ref{WcWtc}) and (\ref{dLaisLa}). We have checked that this match
holds true at the next-order in the $\beta$-expansion of the solutions of the
chiral ring equations. Therefore, thanks to the CS terms also in the 3d/5d case we can realize
the same kind of relation which was manifest in the 2d/4d theories.

In Appendix \ref{CStermsSU4}, we discuss a slightly more complicated example in which a similar phenomenon
occurs. It is a surface operator of type [1,1,2] in the SU(4) gauge theory compactified on $S_\beta^1$.
In this case as well, the two dual quivers lead to the same twisted  chiral superpotential provided 
suitable CS couplings are turned on.

\section{Conclusions and perspectives}
\label{conclusions}

In this paper we have studied surface operators in four- and five-dimensional SU($N$) gauge theories,
focusing on the effective twisted chiral superpotential which governs their infrared dynamics.
Our results are a natural generalization and extension of those in \cite{Gaiotto:2013sma}. 

We have illustrated our findings in the context of the simplest defects in SU(2) and SU(3) theories, even though  
we have performed several checks in a number of theories with higher rank gauge groups. 
Already in the simplest SU(3) case we could observe that different realizations of the same surface 
operator in terms of dual quivers lead, in the five-dimensional case, to different twisted
superpotentials. We have found that this feature is reproduced also in the localization approach where the
different superpotentials arise from different choices of the Jeffrey-Kirwan residue prescription.
In an interesting twist, we have shown that the duality between the coupled 3d/5d quivers can 
be restored by the addition of suitable three-dimensional Chern-Simons terms. Clearly, it would be 
desirable to do 
a systematic analysis of this phenomenon and thoroughly explore the effects of the Chern-Simons
couplings, but we leave this to future work. 

It would be very interesting to extend our results to surface operators in ${\mathcal N}=2^{\star}$ theories. 
As shown in our earlier work \cite{Ashok:2017odt}, the non-perturbative S-duality group of the 
four-dimensional theory constrains the twisted superpotential of the monodromy defect, which can
be written in terms of elliptic and modular forms in a semi-classical expansion. Therefore, it would 
be worthwhile to understand if it is possible to obtain these exact results directly from the chiral ring 
analysis of a coupled quiver gauge theories and also to extend them to five dimensions. 

Surface operators in four- and five-dimensional gauge theories have been studied also by exploiting 
their connections to integrable systems and in particular the relation between the twisted chiral ring 
equations and Bethe ansatz for integrable spin chains \cite{Nekrasov:2009uh, Nekrasov:2009ui, Nekrasov:2009rc}. 
In this context, the wave-functions of the quantum systems can be related to the instanton 
partition function in the presence of surface operators \cite{Kozlowski:2010tv, Nekrasov:2013xda}. 
For the 3d/5d theories studied in this work, recently there has been interesting developments
on the connection between the instanton partition function and the wave functions of relativistic Toda 
theories \cite{Sciarappa:2017hds}. It would be worthwhile to explore this direction using our methods. 
Another interesting possibility is to use dualities between three-dimensional quiver gauge theories with 
flavor to study bi-spectral dualities between quantum integrable systems \cite{Gaiotto:2013bwa}. It would be desirable to investigate the possible implications of our results for these integrable systems, especially in the presence of Chern-Simons couplings. 

\vskip 1.5cm
\noindent {\large {\bf Acknowledgments}}
\vskip 0.2cm
We would like to thank Sourav Ballav, Noppadol Mekareeya, Madhusudhan Raman and Jan Troost for many useful 
discussions. S.K.A. would especially like to thank for the hospitality 
the Physics Department of the University of Torino and the Torino Section of INFN where this work was initiated. 

\noindent
The work of M.B., M.F. and A.L. is partially supported by the MIUR PRIN Contract 
2015MP2CX4 ``Non-perturbative Aspects Of Gauge Theories And Strings''.
\vskip 1cm
\appendix
\section{Chiral correlators in 5d}
\label{app:chiralcorr5d}

In this appendix we briefly review some well-known results about the way in which chiral correlators
are computed using localization \cite{Bruzzo:2002xf,Losev:2003py,Flume:2004rp} that are useful for 
the calculations presented in the main text. 
For details we refer to \cite{Ashok:2016ewb} and references therein.

In a four-dimensional theory SU($N$) the generating function of all chiral correlators of the form
$\left\langle\Tr \Phi^{\ell}\right\rangle$ is
\begin{equation}
\label{mastervev}
\left\langle \Tr \rme^{z\Phi} \right\rangle 
= \sum_{i=1}^N \rme^{ z a_i} - \frac{1}{Z_{\text{inst}}}\sum_{k=1}^\infty  
\frac{q^k}{k!} \int \prod_{I=1}^k \frac{d\chi_I}{2\pi\ii}\, z_k(\chi_I)\, {\mathcal O}(z, \chi_I)
\end{equation}
where $z_k(\chi_I)$ is the $k$-instanton partition function and ${\mathcal O}$ is the following
observable
\begin{equation}
{\mathcal O}(z, \chi_I) = \sum_{I=1}^k \rme^{x\chi_I} (1-\rme^{z \epsilon_1})(1-\rme^{z\epsilon_2}) ~.
\end{equation}
Rather interestingly, the same formula (\ref{mastervev}) can also be exploited to compute the quantum
corrected correlators in the five-dimensional SU($N$) theory provided one uses the appropriate function
$g(x)$ as in (\ref{g5d}) and sets $z= \ell\beta$ for $\ell\in \mathbb{Z}$ and $\ell < N$.  

With an explicit calculation, we find the following universal formula 
\begin{equation}
V_{\ell} \equiv \left\langle \Tr \rme^{\ell \beta \Phi}\right\rangle 
= \sum_{i=1}^N A_{i}^{\ell} + \ell^2 (\beta\Lambda)^{2N}\sum_{i=1}^N 
\frac{A_i^{N-2+\ell}}{\prod_{j\not= i}(A_{i}-A_{j})^2} + O\left((\beta\Lambda)^{4N}\right)
\end{equation}
where $A_i=\rme^{\beta a_i}$. Once the $V_{\ell}$ are obtained, the $U_\ell$'s which appear in 
the five-dimensional Seiberg-Witten curve can be calculated by forming the symmetric polynomials in the usual manner.
In particular, we have $U_1=V_1$ and $U_{N-1}=V_{-1}$. The last relation follows by utilizing the special
unitary condition $\sum_{i=1}^Na_i=0$ which implies $\prod_{i=1}^NA_i=1$. For example, 
for SU(2) we have
\begin{equation}
U_1 = A_1+A_2 + (\beta\Lambda)^4\, \frac{A_1+A_2}{(A_1-A_2)^2} +O( (\beta\Lambda)^{8})~,
\label{U12}
\end{equation}
while for SU(3) we have
\begin{align}
U_1 =&\, A_1+A_2+A_3 + (\beta\Lambda)^6\Big(\frac{A_1^2}{(A_{1}-A_2)^2(A_{1}-A_3)^2}+\notag\\
&+\frac{A_2^2}{(A_{2}-A_1)^2(A_{2}-A_3)^2}+\frac{A_3^2}{(A_{3}-A_1)^2(A_{3}-A_2)^2}\Big)+
O( (\beta\Lambda)^{12})~,
\label{U13}\\
\notag\\
U_2 =&\,A_1A_2+A_2A_3+A_3A_1 
+ (\beta\Lambda)^6\Big(\frac{1}{(A_{1}-A_2)^2(A_{1}-A_3)^2}+\notag\\
&+\frac{1}{(A_{2}-A_1)^2(A_{2}-A_3)^2}+\frac{1}{(A_{3}-A_1)^2(A_{3}-A_2)^2}\Big)+
O( (\beta\Lambda)^{12})~.\label{U23}
\end{align}

\section{Chern-Simons terms in an SU(4) example}
\label{CStermsSU4}

In this section, we provide more evidence towards the duality that was discussed in Section~\ref{secn:dual}. 
We consider the gauge group SU(4) and the surface operator described by the partition [1,1,2]. 
There are two dual descriptions for this defect in terms of quiver diagrams: one is represented
in Fig.~9, and the other is represented in Fig.~10.
In both cases we have added CS interactions. In particular, 
following \cite{Dunne:1998qy, Tong:2000ky, Aganagic:2001uw, Aharony:2014uya}, we have turned 
on a CS terms in those gauge nodes 
where the effective number of fermions is odd, which for both quivers of our example are the U(2) nodes.
One way to justify this is to start from a parity invariant theory and generate these non-integer
CS terms by integrating out an odd number of fermions.
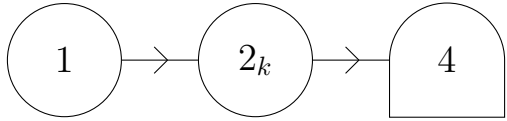
\begin{figure}[ht]
\begin{center}
\begin{tikzpicture}[decoration={
markings,% switch on markings
mark=at position 0.6 with {\draw (-5pt,-5pt) -- (0pt,0pt);
                \draw (-5pt,5pt) -- (0pt,0pt);}}]
  \matrix[row sep=10mm,column sep=5mm] {
      \node(g1)[gauge] {\Large $1$};  & & \node(g2)[gauge] {\Large $2_k$}; 
      & &\node(gfN)[gaugedflavor]{\Large $4$};\\
  };
\graph{(g1) --[postaction={decorate}](g2)--[postaction={decorate}](gfN);};
\end{tikzpicture}
\end{center}
\vspace{-0.5cm}
\caption{The quiver diagram representing the defect SU(4)[1,1,2] with a CS term on the U(2) node.}
\label{quiverSU4}
\end{figure}

\begin{figure}[ht]
\begin{center}
\begin{tikzpicture}[decoration={
markings,% switch on markings
mark=at position 0.6 with {\draw (-5pt,-5pt) -- (0pt,0pt);
                \draw (-5pt,5pt) -- (0pt,0pt);}}]
  \matrix[row sep=10mm,column sep=5mm] {
      \node(gfN)[gaugedflavor] {\Large $4$};  & & \node(g2)[gauge] {\Large $3$};  
      & &\node(g1)[gauge]{\Large $2_{\tilde{k}}$};\\
  };
\graph{(gfN) --[postaction={decorate}](g2)--[postaction={decorate}](g1);};
\end{tikzpicture}
\end{center}
\vspace{-0.5cm}
\caption{The dual quiver diagram representing the defect SU(4)[1,1,2] with a CS term on the U(2) node.}
\label{quiverSU4d}
\end{figure}
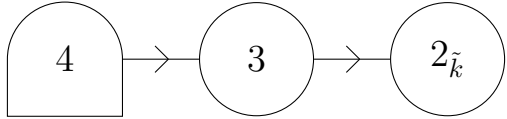

Let us first consider the quiver of Fig.~9. The corresponding twisted effective superpotential with a CS 
coupling $k$ is
\begin{equation}
\begin{aligned}
W&=2\log(\beta\Lambda_1)\,\sigma^{(1)}
+3\log(\beta\Lambda_2)
\sum_{s=1}^2\sigma_s^{(2)}\\
&\qquad+\frac{k}{2}\,\beta\sum_{t=1}^2(\sigma_t^{(2)})^2
-\sum_{t=1}^2\ell(\sigma^{(1)}-\sigma_t^{(2)})-\sum_{t=1}^2
\Big\langle\Tr \ell(\sigma_t^{(2)}-\Phi)\Big\rangle~.
\end{aligned}
\end{equation}
The twisted chiral ring relation at the first node is
\begin{align}
\widehat{Q}_2(\sigma^{(1)})-\left(\beta\Lambda_1\right)^2=0~,
\end{align} 
while at the second node we have
\begin{equation}
\widehat{P}_4(\sigma_s^{(2)})=-\left(\beta^3\Lambda_2^3\,
\widehat{Q}_1(\sigma_s^{(2)})\,(S_s^{(2)})^k
+\frac{\beta^5\Lambda^8}{\Lambda_2^3 \widehat{Q}_1(\sigma_s^{(2)})\,(S_s^{(2)})^k}\right)
\end{equation}
for $s=1,2$. Solving these equations order by order in $\beta$ by using the standard ansatz and
the chiral correlators of the SU(4) theory, we obtain
\begin{equation}
\begin{aligned}
\sigma^{(1)}_\star&=a_1+\beta\,\frac{A_1^{1/2}A_2^{1/2}}{A_{12}}\,\Lambda_1^2+\beta^2\,
\frac{A_1^{1-k}A_3^{1/2}A_4^{1/2}}{A_{13}A_{14}}\,\frac{\Lambda^8}{\Lambda_1^2\Lambda_2^3}
+\ldots~,\phantom{\bigg|}\\
\Tr \sigma^{(2)}_{\star}&=a_1+a_2-\beta^2\,\left(
\frac{A_2^{1+k}A_3^{1/2}A_4^{1/2}}{A_{23}A_{24}}\,\Lambda_2^3
-\frac{A_1^{1-k}A_3^{1/2}A_4^{1/2}}{A_{13}A_{14}}\,\frac{\Lambda^8}{\Lambda_1^2\Lambda_2^3}\right)+\ldots~.
\end{aligned}
\label{sol}
\end{equation}

We now consider the dual quiver represented in Fig.~10. In this case the twisted superpotential with a
CS coupling $\widetilde{k}$ is
\begin{equation}
\begin{aligned}
\widetilde W&=-2\log(\beta\widetilde{\Lambda}_1)
\sum_{s=1}^3\widetilde\sigma_s^{(1)}-3\log(\beta\widetilde{\Lambda}_2)
\sum_{t=1}^2\widetilde\sigma_t^{(2)}
+\frac{\widetilde k}{2}\,\beta\sum_{t=1}^2(\widetilde\sigma_t^{(2)})^2\\
&\qquad
-\sum_{s=1}^3\ell(\widetilde\sigma_s^{(1)}-\widetilde\sigma_1^{(2)})
-\sum_{s=1}^3\ell(\widetilde\sigma_s^{(1)}-\widetilde\sigma_2^{(2)})-\sum_{s=1}^3\Big\langle\Tr \ell(\Phi-\widetilde\sigma_s^{(1)})\Big\rangle~.
\end{aligned}
\end{equation}
The corresponding chiral ring equations are
\begin{equation}
\widehat{P}_4(\widetilde\sigma_s^{(1)})=\left(\beta^2\,
\widehat{Q}_2(\widetilde\sigma_s^{(1)})\,\widetilde{\Lambda}_1^2+\beta^6\,
\frac{\Lambda^8}{\widehat{Q}_2(\widetilde\sigma_s^{(1)})\,\widetilde{\Lambda}_1^2}\right)
\end{equation}
for $s=1,2,3$, and
\begin{equation}
\widehat{Q}_1(\widetilde\sigma_t^{(2)})=-\beta^3
\frac{\widetilde{\Lambda}_2^3}{(\widetilde S_t^{(2)})^{\widetilde k}}
\end{equation}
for $t=1,2$.
Solving these equations with the usual ansatz, we find
\begin{align}
\Tr \widetilde{\sigma}^{(1)}_{\star}&=a_2+a_3+a_4-\beta\,\frac{A_1^{1/2}A_2^{1/2}}{A_{12}}
\,\widetilde{\Lambda}_1^2\cr
&\qquad+\beta^2\,
\bigg(\frac{A_1^{1/2}A_3^{1+\widetilde k}A_4^{1/2}}{A_{31}A_{34}}+\frac{
A_1^{-\widetilde{k}-1/2}A_2^{-\widetilde k-1}A_3^{-\widetilde{k}-1/2}}{A_{41}A_{43}}\bigg)
\frac{\Lambda^8}{\widetilde{\Lambda}_1^2\widetilde{\Lambda}_2^3}+\ldots~,\notag\\
\label{soldual}\\
\Tr \widetilde{\sigma}^{(2)}_{*}&=a_3+a_4-
\beta^2\bigg(\frac{A_2^{1/2}A_3^{1-\widetilde{k}}A_4^{1/2}}{A_{32}A_{34}}
+\frac{A_1^{\widetilde{k}-1}A_2^{-1/2+\widetilde k}
A_3^{-1/2+\widetilde k}}{A_{42}A_{43}}\bigg)\widetilde{\Lambda}_2^3\notag\\
&\qquad+\beta^2\bigg(\frac{A_1^{1/2}A_3^{1+\widetilde k}A_4^{1/2}}{A_{31}A_{34}}
+\frac{A_1^{-1/2-\widetilde k}A_2^{-\widetilde k-1}A_3^{-1/2-\widetilde k}}{A_{14}A_{34}}\bigg)
\frac{\Lambda^8}{\widetilde{\Lambda}_1^2\widetilde{\Lambda}_2^3}+\ldots~.
\notag
\end{align}
These expressions look very different from the solution (\ref{sol}) of the chiral ring equations of the original
quiver. However, if we impose the SU(4) tracelessness constraint $\sum_ia_i=0$ 
and use the following map
\begin{align}
k=-\widetilde k=\frac{1}{2}\qquad\text{and}\qquad \Lambda_I ={\widetilde \Lambda}_I~,
\end{align}
we find the following relations
\begin{equation}
\begin{aligned}
\Tr \widetilde\sigma^{(1)}_{\star}&=-\sigma^{(1)}_{\star}~,\\
\Tr \widetilde\sigma^{(2)}_{\star}&=-\Tr \sigma^{(2)}_{\star} ~.
\end{aligned}
\end{equation}
This proves that, to leading order in the instanton expansion, 
the superpotentials of the dual pair match as expected. We have checked that 
this match continues to hold up to two instantons as well.

\providecommand{\href}[2]{#2}\begingroup\raggedright

\endgroup

\end{document}